\documentclass[longauth]{aa}  
\usepackage{graphicx}
\usepackage{txfonts}
\usepackage{hyperref}
\hypersetup{
    colorlinks=true,
    linkcolor=blue,
    filecolor=blue,      
    urlcolor=blue,
    citecolor=blue}

\begin{document} 
   \title{VVV-WIT-13: an eruptive young star with cool molecular features}
   \author{Zhen Guo \inst{1,2}\fnmsep\thanks{zhen.guo@uv.cl}
          \and
          P. Lucas\inst{3}
          \and
          S. N. Yurchenko\inst{4}
          \and
          T. Kaminski\inst{5}
          \and
          M. Montesinos\inst{6}
          \and
          S. Nayakshin\inst{7}
          \and
          V. Elbakyan\inst{8,9}
          \and         
          J. Osses\inst{1,2}
          \and
          A. Caratti~o~Garatti\inst{10} 
          \and
          H. Zhao\inst{11}
          \and  
          R. Kurtev\inst{1,2}
          \and
          J. Borissova\inst{1,2} 
          \and
          C. Morris\inst{1,3} 
          \and
          D. Minniti\inst{11,12}
          \and
          J. Alonso-García\inst{13,2}
          \and
          V. Fermiano \inst{14}
          \and
          R.~K.~Saito \inst{14}
          \and
          N.~Miller \inst{15}
          \and
          G.~Zsidi \inst{3}
          \and
          H.~D.~S.~Muthu \inst{3}
          \and
          C.~Briceño\inst{16}  
          \and
          C.~Contreras~Peña\inst{17, 18}  
          \and
          A. E. Lynas-Gray\inst{4, 19, 20}
          \and
          J.~Tennyson\inst{4} 
          \and
          L.~Wang\inst{21, 22}
          \and
          L.~Yu\inst{22}
          \and
          D.~Benitez-Palacios\inst{1}
          \and
          J.~Yang\inst{23}
          \and
           M.~Kuhn\inst{3} 
          \and
          S.~X.~Wang\inst{24}
          }
   \institute{Instituto de F{\'i}sica y Astronom{\'i}a, Universidad de Valpara{\'i}so, ave. Gran Breta{\~n}a, 1111, Casilla 5030, Valpara{\'i}so, Chile\\
              \email{zhen.guo@uv.cl}
         \and
         Millennium Institute of Astrophysics, Nuncio Monse{\~n}or Sotero Sanz 100, Of. 104, Providencia, Santiago, Chile
          \and
          Centre for Astrophysics Research, University of Hertfordshire, Hatfield AL10 9AB, UK
          \and
        Department of Physics and Astronomy, University College London, Gower Street, London WC1E 6BT, UK
        \and
        Nicolaus Copernicus Astronomical Center, Polish Academy of Sciences, Rabianska 8, 87-100 Toru\'n, Poland
         \and
          Departamento de Física, Universidad Técnica Federico Santa María, Avenida España 1680, Valparaíso, Chile
         \and
        School of Physics and Astronomy, University of Leicester, Leicester LE1 7RH UK
        \and
        Fakultat fur Physik, Universitat Duisburg-Essen, Lotharstraße 1, D-47057 Duisburg, Germany
        \and
        Research Institute of Physics, Southern Federal University, Rostov-on-Don 344090, Russia
        \and
         INAF - Osservatorio Astronomico di Capodimonte, salita Moiariello 16, 80131, Napoli, Italy
        \and
       Departamento de Fisica y Astronomia, Facultad de Ciencias Exactas, Universidad Andres Bello, Fernandez Concha 700, 8320000 Santiago, Chile
        \and
        Vatican Observatory, V00120 Vatican City State, Italy
        \and
        Centro de Astronom{\'i}a (CITEVA), Universidad de Antofagasta, Av. Angamos 601, 02800 Antofagasta, Chile
        \and
        Departamento de F{\'i}sica, Universidade Federal de Santa Catarina, Trindade 88040-900, Florianopol{\'i}s, SC, Brazil
        \and
        University of Wyoming, 1000 E University Ave, Laramie, WY USA
        \and
        Cerro Tololo Inter-American Observatory, National Optical Astronomical Observatory, Casilla 603, La Serena, Chile
        \and
        Department of Physics and Astronomy, Seoul National University, 1 Gwanak-ro, Gwanak-gu, Seoul 08826, Republic of Korea
        \and
        Research Institute of Basic Sciences, Seoul National University, Seoul 08826, Republic of Korea
        \and
        Department of Physics, University of Oxford, Keble Road, Oxford OX1 3RH, United Kingdom
        \and
        Department of Physics and Astronomy, University of the Western Cape, Bellville 7535, South Africa
        \and
        Chinese Academy of Sciences South America Center for Astronomy (CASSACA), National Astronomical Observatories, CAS, Beijing 100101, China
        \and
        Departamento de Astronom{\'i}a, Universidad de Chile, Las Condes, 7591245 Santiago, Chile
        \and
        Department of Astronomy, the University of Michigan, 1085 S. University, 323 West Hall, Ann Arbor, MI 48109-1107, USA
       \and
       Department of Astronomy, Tsinghua University, Beijing 100084, People’s Republic of China
       }

   \date{Received xxx; accepted xxx}

  \abstract
   {Outburst phenomena are observed at different stages of stellar evolution, due to the enhancement of mass accretion rate on protostars or even stellar merger events.  {In the case of a Young Stellar Object (YSO), the episodic mass accretion event plays an important role in the pre-main-sequence stellar mass assembly.} Here we investigate an infrared eruptive source (RA = 16:53:44.38; DEC = -43:28:19.47), identified from the decade-long VISTA Variables in the V\'ia L\'actea survey (VVV). We named this target after a group of variable sources discovered by VVV, as VVV-WIT-13, with WIT standing for "What Is This?", due to its unique photometric variation behaviour and the mysterious origin of the outburst. This target exhibited an outburst with a 5.7~mag amplitude in the $K_s$-band, remained on its brightness plateau for 3.5 years, and then rapidly faded to its pre-eruptive brightness afterwards.}
   {We aim to reveal the variable nature and outburst origin of VVV-WIT-13 by presenting our follow-up photometric and spectroscopic observations along with theoretical models.}
   {We gathered photometric time series in both near- and mid-infrared wavelengths. We obtained near-infrared spectra during the outburst and decaying stages on XSHOOTER/VLT and FIRE/Magellan, and then fitted the detected molecular absorption features using models from ExoMol. We applied 2D numerical simulations to re-create the observables of the eruptive phenomenon.}
   {We observe deep AlO absorption bands in the infrared spectra of VVV-WIT-13, during the outburst stage, along with other more common absorption bands (e.g. CO). Our best-fit model suggests a 600~K temperature of the AlO absorption band. In the decaying stage, the AlO bands disappeared, whilst broad blue-shifted  H$_2$ lines arose, a common indicator of stellar wind and outflow. The observational evidence suggests that the CO and TiO features originate from an outflow or a wind environment.}
   {We find that VVV-WIT-13 is an eruptive young star with instability occurring in the accretion disk. One favoured theoretical explanation of this event is a disrupted gas clump at a distance of 3~au from the source. If confirmed, this would be the first such event observed in real time.}

   \maketitle

\section{Introduction}

In time-domain astronomy, bursts are a common type of transient phenomenon. They are often associated with sudden increases in mass accretion rate, as observed in young stellar objects (YSOs) and active galactic nuclei \citep{Scaringi2015}, or with close interactions between two astrophysical bodies, such as in classical and dwarf novae \citep{Poggiani2024}, tidal disruption events \citep{Rees1988}, or planet engulfment scenarios \citep{De2023}. In particular, photometric surveys have discovered luminosity spreads among young clusters, which are a consequence of the episodic accretion scenario \citep{Fischer2023}.  {During the episodic accretion stage, the stellar mass accretion rate can increase by orders of magnitude.} The duration and frequency of these high-accretion phases play a crucial role in stellar mass assembly. For instance, during the century-long outburst of FU~Ori, the mass accretion rate has been enhanced by 4-5 orders of magnitude \citep{Zhu2007}.
During these accretion bursts, the inner accretion disks are thoroughly heated, causing the snowlines of molecular species to move outwards \citep{Abraham2009}. This process creates opportunities for the growth of planetesimals and affects the formation of young planets \citep{Jorgensen2020}. Questions have been raised about the triggering mechanisms of the outbursts around young stars, particularly whether they can only happen under specific conditions, such as gravitational perturbations from a young giant planet.
 
Various physical mechanisms have been proposed to trigger accretion bursts on young stars, such as magneto-rotational instability \citep[MRI;][]{Armitage2001, Zhu2009b, Elbakyan2021} and thermal instability \citep[TI;][]{Lodato2004, Nayakshin2024b}. Moreover, perturbations from a secondary body in the stellar system may also lead to accretion bursts, including the gravitational instability (GI) introduced by a stellar flyby \citep{Cuello2019, Borchert2022} and an evaporating giant planet located in a close-in orbit \citep{Nayakshin2024}. Alternatively, in more evolved stellar systems without accretion disks, planet engulfment events can trigger a short-lived optical outburst \citep[e.g.][]{Stephan2020, De2023, Soker2023}, while the merger of stellar binaries results in red novae that last for months to years and at luminosities $\lesssim 10^6$ L$_\odot$ \citep{Soker2006, Kaminski2018}.

In this work, we present photometric and spectroscopic analyses of the infrared eruptive object VVV-WIT-13 (RA = 16:53:44.38; DEC = -43:28:19.47). The acronym WIT stands for "What Is This?", indicating the mysterious nature of this source. The designation 13 belongs to a list of intriguing variables \citep[e.g.][]{Minniti2017, Lucas2020WIT, Smith2021, Saito2023}, discovered by the infrared VISTA Variable in the Via Lactea survey \citep[VVV;][]{Minniti2010}.  As with our previous discoveries, VVV-WIT-13 represents a challenging astrophysical problem. A few years passed after the discovery until we were able to gather enough ancillary data (in this case, additional photometry and spectroscopy) in order to be able to discard some of the varied scenarios and explanations that were initially proposed. The infrared variability of VVV-WIT-13 was originally reported in \citet{Contreras2017}, as source VVVv746 with $\Delta K_s =$ 1.27~mag between 2010 and 2014. \citet{Lucas2020} reported a $\Delta W1 = 5.26$~mag outburst in this object in the mid-infrared WISE bands, under the name WISEA J165344.39-432819.2, classifying it as an eruptive young stellar object. More recently, this object was presented in a collection of VVV high-amplitude variable sources (with a burst of $\Delta K_s = 5.67$ mag) as L222$\_$59 \citep{Lucas2024}. Our follow-up spectroscopic study discovered deep AlO absorption features in the near-infrared spectra of this source taken during the outburst stage \citep{Guo2024a}, which have previously only been observed among the spectra of red nova outbursts and Miras \citep{Banerjee2012, Kaminski2015, Steinmetz2024}. Based on this, \citet{Guo2024a} classified L222$\_$59 as a post-main-sequence nova-like object. However, this classification conflicts with the young characteristics of this source observed during the quiescent stage, prompting this paper to investigate the true nature of this unique eruptive object.

 The VVV-WIT-13 is projected within the Galactic disk, associated with several star-forming regions visible in \textit{Spitzer} images (see Figure~\ref{fig: Spitzer_map}). VVV-WIT-13 is located $6\arcsec$ away from the Galactic molecular cloud SDG G342.136+0.2045 \citep{Duarte2021}. This molecular cloud has an approximate distance of 3.2 $\pm$ 0.5 kpc to the solar system, estimated by the radial velocity of $^{13}$CO (2-1) emission. The far-infrared emission of this cloud was detected by Herschel \citep{Peretto2016}, which probably outshines the emission from the disk/envelope of VVV-WIT-13.
 Additionally, within a 5$\arcmin$ radius of VVV-WIT-13, there are five \textit{Herschel} Hi-GAL clumps, with the nearest being only 17\arcsec away \citep[HIGALBM 342.1366+0.2006;][]{Elia2017}. Based on a 2~kpc distance \citep[see][]{Mege2021}, \citet{Elia2017} fitted a mass of $40\, M_\odot$ for the nearest cold core. VVV-WIT-13 is also surrounded by several Galactic H {\sc ii} regions, such as GAL 342.09+00.42, G341.97+00.44 and G342.36+00.11 (see Figure~\ref{fig: Spitzer_map}). Based on our previous statistical studies, VVV-WIT-13 is likely a young star given the large number of star-formation-related sources within a $5\arcmin$ radius \citep[see Figure 8 in][]{Contreras2017}. However, VVV-WIT-13 lacks any optical counterparts in existing photometric catalogues because it resides within a \textit{Spitzer} dark cloud \citep[G342.135+0.204;][]{Peretto2009}. According to previous investigations from the VVV survey, such an overlap has a chance of only 1\% in the southern Galactic plane \citep{Lucas2020WIT} and therefore VVV-WIT-13 is liked to be associated with the IR dark cloud.

 VVV-WIT-13 is classified as a candidate YSO by its infrared colour indices from the \textit{Spitzer} GLIMPSE survey, where it was originally named  [RMB2008] G342.1371+00.2054 \citep{Robitaille2008}. In the SPICY YSO catalogue \citep[][SPICY ID 43459]{Kuhn2021}, it is identified as a flat-spectrum YSO, representing an intermediate stage between a protostar and a T Tauri star. The classification was based on the variability and the {\it Spitzer} infrared photometry taken in 2006. VVV-WIT-13 belongs to a spatial association of YSOs (SPICY G342.1+0.2) with 442 member sources along the Galactic disk plane. In this group, 91 members have counterparts (spatial separation less than 1$\arcsec$) in the \textit{Gaia} DR3 data release \citep{Gaiadr3}. { After removing outliers in the proper motion and parallax (see Section \ref{sec:discussion_group}), we found a median parallax of 0.487 mas, corresponding to a distance of 2.05~kpc.} The 2~kpc median distance is consistent with the distance of the nearby gas clump HIGALBM 342.1366+0.2006 \citep{Mege2021}. The kinematic velocity of VVV-WIT-13 is expected to be $V_{\rm LSR}$=-22~km/s \citep{Wenger2018} given the distance and Galactic location. Unless otherwise stated, we assume a distance of 2~kpc to the VVV-WIT-13.


\section{Observational data}
\label{sec:obs}
\subsection{Photometric data and catalogues}
  \begin{figure*}
   \centering
   \includegraphics[width=18cm]{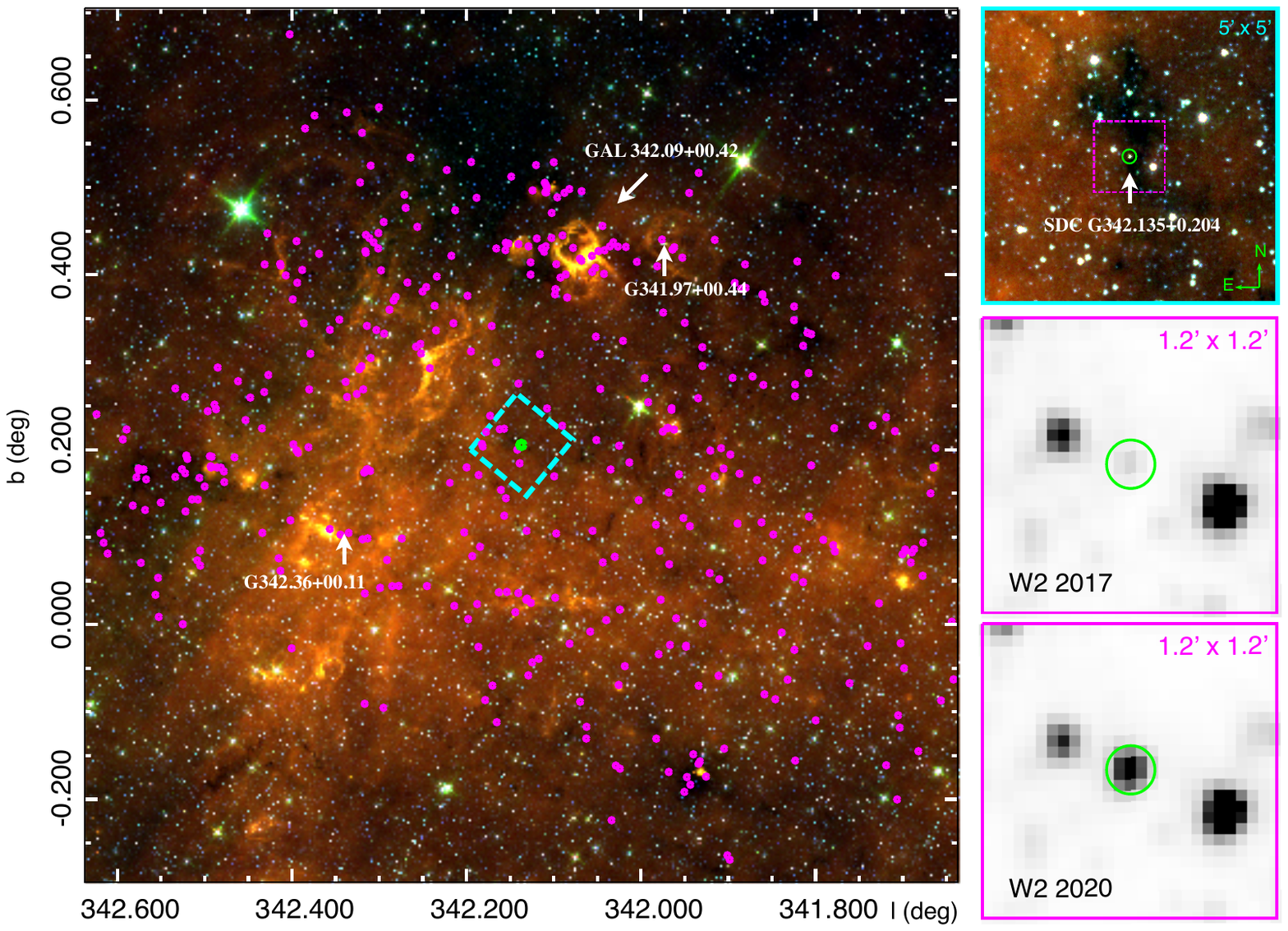}
      \caption{Infrared images of the surrounding environment of VVV-WIT-13 (located in the center and marked by the green circle in all frames). {\it Left:} {\it Spitzer IRAC} three-colour image (1 $\times$ 1 deg, blue: [4.5], green: [5.8], red: [8.0]), taken in 2004 during the pre-outburst stage, with Galactic longitude (l) and latitude (b). Sources belonging to the G342.1+0.2 YSO group, selected by {\it SPICY}, are marked by magenta circles. {\it Upper right:} a zoomed-in 5$\arcmin$ $\times$ 5$\arcmin$ {\it Spitzer IRAC} three-colour image, with the equatorial North-up orientation.  {\it Middle and bottom right:} 1.2$\arcmin$ $\times$ 1.2$\arcmin$ $W2$-band images from {\it unWISE} \citep{Lang2014a, Meisner2017}, taken in 2017 and 2020.}
    \label{fig: Spitzer_map}
   \end{figure*}

\subsubsection{The near-IR photometry}
The VISTA Variables in Via Lactea survey (VVV) is a public survey of the European Southern Observatory (ESO), which obtained near-infrared photometric light curves of millions of sources towards the inner Galactic disk plane and the Galactic Centre from 2010 to 2016. The vast majority of the time series observations were taken in the $K_s$-band (2.15~$\mu$m), with a few multi-colour visits \citep{Minniti2010}. Starting in 2016, the VVV eXtended survey (VVVX), as an extension of VVV, further obtained a dozen epochs of multi-colour photometry in the near-infrared, with twice as large sky coverage as the original VVV field \citep[see][]{Minniti2016, Saito2024}. We obtained the VVV photometric measurements from the VIRAC2-$\beta$ catalogue for detections in $H$ and $K_s$ bands \citep{Smith2025}. We resampled the $K_s$ photometry to 1-day bins to increase the photometric accuracy. After removing outliers beyond 3$\sigma$ of each bin, the final light curve comprises the median values from each bin. We also performed aperture photometry on $Z$, $Y$ and $J$-band VVV images. The photometric measurements in $Y$ (2011 epoch) and $J$-bands are listed in Table~\ref{tab:photometry}. 

We conducted two photometric observations on the ESO New Technology Telescope, using the Son OF ISAAC (SOFI) imager \citep{Moorwood1998}, in May 2021 and April 2022. In 2023, 2024 and 2025, we obtained photometric observations on the InfraRed Survey Facility (IRSF) telescope located at the South African Astronomical Observatory \citep{Kato2007}. In June 2025, we observed a single-epoch $K_s$ photometry with the Rapid Eye Mount (REM) telescope at La Silla observatory, with a total exposure time of $3 \times 5 \times 15$s = 225 s. Moreover, we used the $J$-band acquisition image from the Magellan FIRE spectrograph as a photometric data point. Additionally, we conducted aperture photometry on the $K_s$-band image from the 2MASS survey extracted from the NASA/IPAC Infrared Science Archive. 

We reduced images by using custom-written pipelines. In each filter, the frames taken in each dithering cycle were shifted and co-added as a single image. Nearby field stars were selected as photometric references based on their brightness in each band and $K_s$-band stability from the VVV time series ($\delta K_s < 0.05$ mag, standard deviation). Then, the relative magnitude of the target was extracted by aperture photometry methods similar to those in \citet{Guo2018a}. The measured brightness of our target is listed in Table~\ref{tab:photometry}.
     \begin{figure*}
   \centering
    \includegraphics[height=8cm]{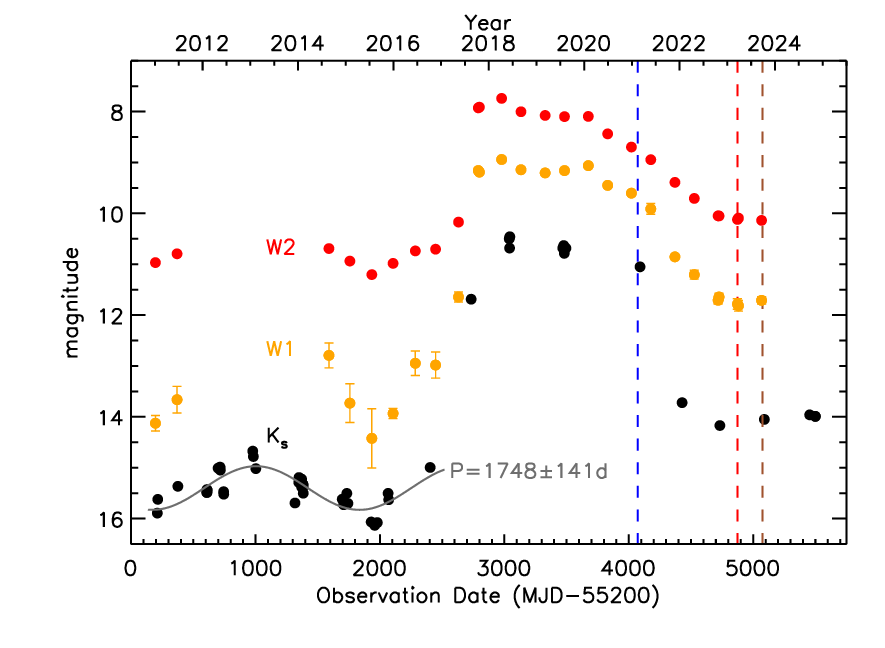}
   \includegraphics[height=8cm]{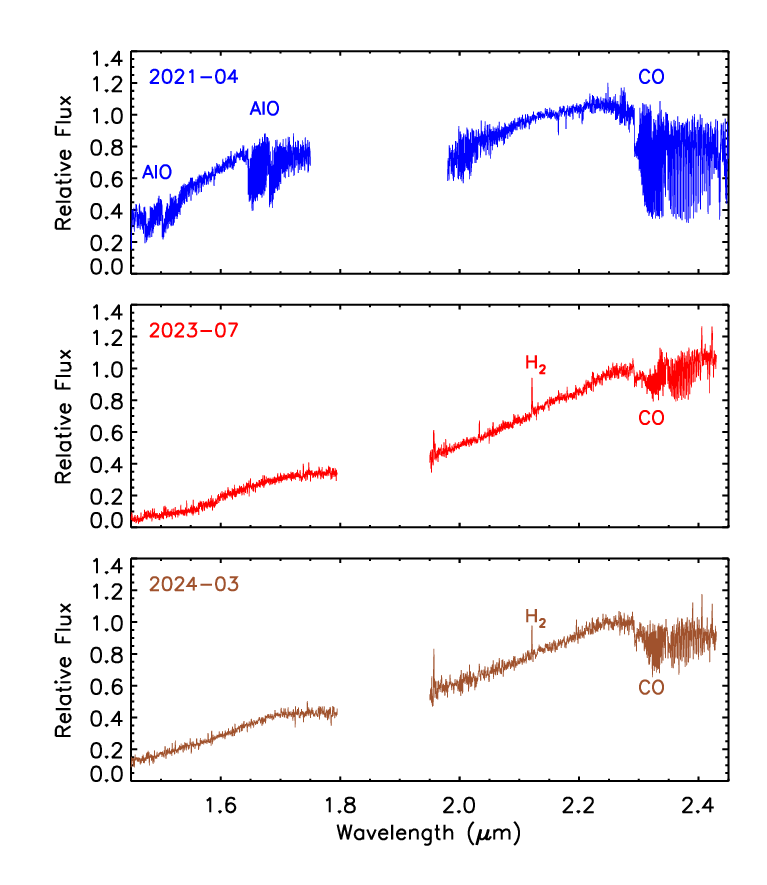}
      \caption{\textit{Left}: infrared light curves ($K_s$, $W1$ and $W2$-band) of VVV-WIT-13. The $W1$-band error bars (orange) are the standard deviation of the magnitude spread in each epoch before binning. The typical error bar in the $K_s$ band ($<$ 0.1 mag) is smaller than the symbol size. A sinusoidal fitting result is presented by the grey line. The spectroscopic epochs are marked by the vertical dashed lines with the same colour scheme as the right panels. \textit{Right}: near-infrared spectra of VVV-WIT-13 obtained in 2021, 2023 and 2024. Key spectral features are labelled on the plot.}
         \label{fig: spec+lc}
   \end{figure*}
   
\subsubsection{The archival infrared photometry}

The pre-outburst infrared brightness of VVV-WIT-13 was measured by the {\it Spitzer} GLIMPSE survey \citep[][]{Benjamin2003} and the 24~$\mu$m band in MIPSGAL \citep{Carey2009}. It was also detected in {\it ALLWISE} \citep{Wright2010} and {\it NEOWISE} \citep{Mainzer2014} surveys. We downloaded mid-infrared photometric data from the NASA/IPAC Infrared Science Archive (IRSA\footnote{\url{https://irsa.ipac.caltech.edu/}}). Specifically, we obtained single-epoch {\it WISE} photometry and then averaged detections within 3-day bins to increase the accuracy. At the fainter end of the $W1$-band time series ($>$ 12 mag), the intrinsic spread of individual magnitude measurements within one bin becomes larger than the error. Therefore, we used the standard deviation within one bin ($\sigma$) as the error of each binned epoch. Additionally, data points located outside 3-$\sigma$ of each bin were eliminated before calculating the average.  The \textit{unWISE} images shown in Figure \ref{fig: Spitzer_map} are downloaded from the {\it wiseview} online tool\footnote{\url{http://byw.tools/wiseview}}. There is no Herschel Hi-Gal detection of this source, likely shadowed by the nearby gas clump HIGALBM 342.1366+0.2006 \citep{Elia2021}.

\subsection{Spectroscopic observations and data reduction}

We acquired four near-infrared spectra of VVV-WIT-13, including two epochs in 2021 during the brightness plateau, one spectrum taken in 2023 and a final epoch obtained in early 2024. The epochs are marked on the light curve in Figure \ref{fig: spec+lc}. 

On 29th April and 3rd May 2021, we obtained two optical to near-infrared spectra of VVV-WIT-13 using the XSHOOTER spectrograph mounted on the European Southern Observatory Very Large Telescope, under the service mode (PI: Lucas). The VIS and NIR arms of XSHOOTER were used, with a 0.9$"$ slit width in the optical (R = 8900) and a 0.6$"$  slit in the NIR arm (R = 8100). The spectra were extracted by the ESO pipeline with heliocentric correction \citep{Vernet2011, Freudling2013} and the telluric absorption features were corrected by the {\sc molecfit} software \citep{Smette2015, Kausch2015}. We obtained rough flux calibration using the SOFI photometry taken on 7th May 2021, assuming no short-timescale variability. These two spectra were originally published in \citet{Guo2024a} as part of the spectroscopic follow-up campaign of a group of high-amplitude variable sources. In this work, we combined these two spectra to increase the signal-to-noise ratio, as no variability was found between the two epochs.

 A near-infrared spectrum was taken during the fading stage of this source in July 2023 (PI: Guo) with the FIRE spectrograph \citep{Simcoe2013} on the Magellan Baade Telescope at Las Campanas Observatory. We used the 0.6$"$ slit (R = 6000) with a wavelength range between 0.8 to 2.5~$\mu$m and a spatial resolution of 0.18 arcsec per pixel. An A0V telluric standard star with the same airmass was observed after the scientific exposures for telluric correction. The FIRE data was reduced using the {\sc firehose v2.0} pipeline \citep{gagne2015}. Additional custom-designed programs were applied to obtain accurate wavelength calibrations using OH skylines and to fix the discontinuity between the spectral orders. Finally, only $H-$ and $K$-bandpass spectra are presented in this work (1.45 to 2.45~$\mu$m), as the $J$ spectra have a low signal-to-noise ratio (SNR $<$ 2).  See details of our spectroscopic data reduction methods in \citet{Guo2020, Guo2021}.

 Another XSHOOTER spectrum was obtained using the ESO Director's Discretionary Time on 16th March 2024 (PI: Kaminski). The same data reduction was applied as for the earlier Xshooter spectra. The spectra of VVV-WIT-13 are presented in Figure \ref{fig: spec+lc} with spectral epochs marked on the light curves. 

\section{Photometric results}
\label{sec:result}

In this section, we present photometric results of VVV-WIT-13, including its infrared colour indices, the spectral energy distribution and its photometric variability. In most parts of this section, we assumed a young nature of VVV-WIT-13 as indicated by its spatial location.

\subsection{Infrared colour indices of VVV-WIT-13}

\begin{figure}
  \centering
     \includegraphics[height=7cm]{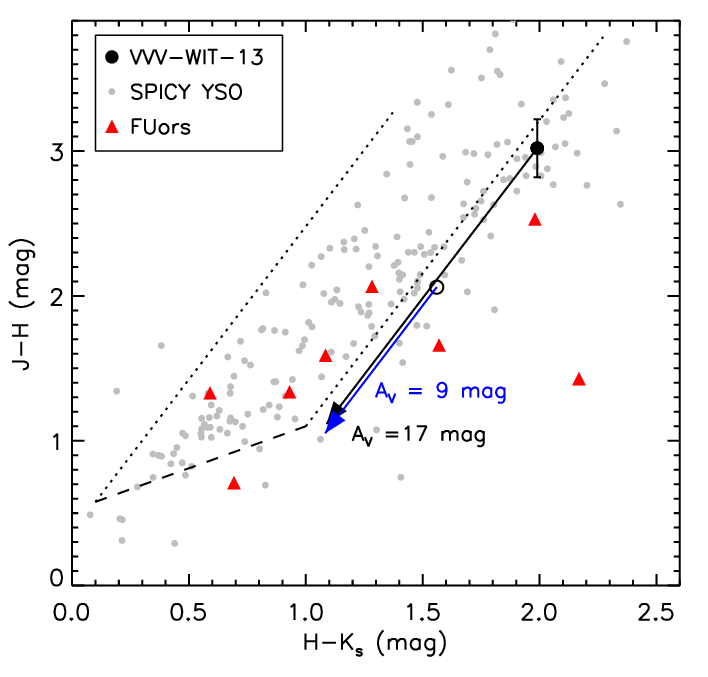}
     \caption{Near-infrared colour-colour diagram of YSOs from the VVV survey. VVV-WIT-13 is marked by the black circles (filled: quiescent; open: outburst). Other members in the SPICY group G342.1+0.2 are presented by the grey dots. A selection of VVV FUor-type eruptive sources is shown in red. The dashed line represents the synthetic locus of Class II YSOs \citep{Meyer1997}. The dotted and solid lines are extinction vectors from \citet{WangS2019}.}
\label{fig: CCD}
\end{figure}

Infrared colours provide essential information on stars, such as the evolutionary stages and line-of-sight extinction in the case of YSOs \cite[][]{Lada1987}. Here, we measured the quiescent and outburst colours of VVV-WIT-13 using the photometric data and compared them with catalogued YSOs. The quiescent near-infrared colours of VVV-WIT-13 ($J-H =$ 3.02 mag; $ H-K_S =$ 1.99 mag) indicate that it is an embedded YSO, { located between Class I and II stages} (see Figure \ref{fig: CCD}). This is consistent with the classification from the SPICY catalogue based on the mid-infrared \textit{Spitzer} Spectral Energy Distribution (SED). We estimated the extinction of VVV-WIT-13 as $A_V \sim 17$ mag, assuming that the de-reddened quiescent colour indices agree with normal disk-bearing YSOs \citep{Meyer1997} and using the extinction law from \citet{WangS2019}. For comparisons, in Figure \ref{fig: CCD} we present the infrared colours of members of the SPICY group G342.1+0.2 and FUor-type objects published in \citet{Guo2024a}. The majority of the YSO candidates have similar colours to reddened Class I/II YSOs.

The near-infrared colours of VVV-WIT-13 during the outburst are $J-H =$ 2.06~mag and $H-K_s =$ 1.56~mag. Both indices are bluer than the pre-outburst quiescent stage, in line with the brighter-and-bluer trend of eruptive YSOs. On the colour-colour diagram, we noticed a coincidence between the colour change of this outburst and the extinction vector. However, the $H$ and $K_s$ band amplitudes ($\Delta H = 5.28$~mag and $\Delta K_s = 4.85$ mag) do not follow the extinction law ($A_H = 1.67 A_{K_s}$).

The classification of YSOs based on infrared colours is often contaminated by the post-main-sequence asymptotic giant branch (AGB) stars. To separate YSOs and post-main-sequence sources, \citet{Robitaille2008} developed empirical cuts on the {\it Spitzer} colours\footnote{YSOs have \textit{IRAC2} $>$ 7.8~mag and \textit{IRAC4} - [24.0] $\geq$ 2.5 mag.}. The \textit{Spitzer} colours of VVV-WIT-13 agree with YSO instead of the AGB identification. In previous publications, we distinguished YSOs and Miras based on their infrared variability and drew regions of Miras/YSOs on the {\it WISE} colour-colour and colour-magnitude diagrams \citep{Lucas2017, Guo2022}.  On the \textit{WISE} colour-colour diagram, the quiescent colour of VVV-WIT-13 is located beyond the region occupied by AGB stars and near the boundary between Miras and YSOs. We examined the periodicity of the $K_s$-band light curves of group members of SPICY G342.1+0.2 and discovered eight Mira candidates. All eight Mira candidates have different locations on the \textit{WISE} colour-colour diagram than VVV-WIT-13 (see Figure \ref{fig: WISE_CCD} in Appendix).

\subsection{The Spectral Energy Distribution of VVV-WIT-13}

 {The infrared SED is a common tool for identifying young stars with a circumstellar disk or envelope. Sources at earlier evolutionary stages (i.e., protostars) exhibit excess emission at longer wavelengths.} A combination of high line-of-sight extinction and thermal emission from the accretion disk/envelope contributes to the infrared excess \citep{Lada1987}.  {In this work, we used the pre-outburst/quiescent near- to mid-infrared SED (between 0.8 to 24~$\mu$m) to classify VVV-WIT-13 and to estimate its bolometric luminosity. The following discussions assume that VVV-WIT-13 is a young star undergoing an outburst since 2018.} The quiescent SED is composed of $J$, $H$ and $K_s$ VVV photometry obtained in the year 2010, \textit{ALLWISE} photometry taken in 2010 and \textit{Spitzer} data observed in 2004. { The outburst SED includes photometric measurements from SOFI and \textit{NEOWISE}. We also added VISTA $Z$- and $Y$-band flux via integrating the XSHOOTER spectrum through the filter transmission curves \citep{Rodrigo2012}. }

\begin{figure*}
  \centering
   \includegraphics[width=7.6cm]{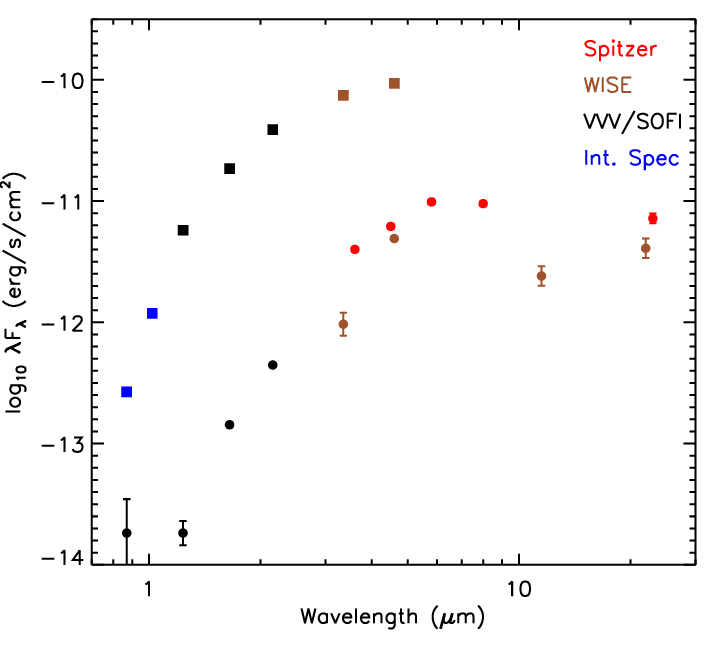}
      \includegraphics[width=7.9cm]{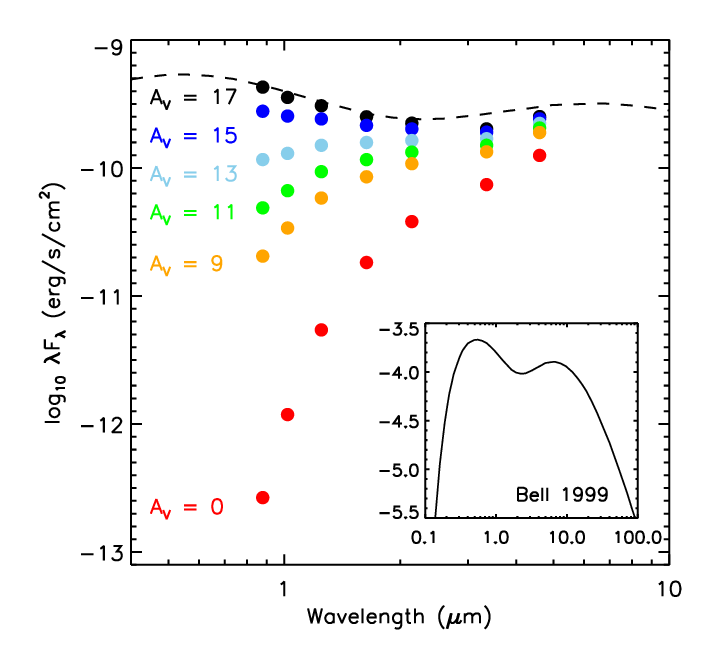}
      \caption{\textit{Left:} Spectral energy distribution (SED) of VVV-WIT-13 with data obtained from VVV, SOFI, {\it Spitzer}, {\it WISE} and integrated spectrum. The pre-outburst SED is presented by dots, and the in-outburst SED is shown by squares.  \textit{Right:} In-outburst SED of VVV-WIT-13 dereddened with $A_V$ ranging from 0 to 17 mag. As a comparison, we present the ``high mass accretion" SED calculated by \citet[][]{Bell1999}. We scaled the model to match the SED of VVV-WIT-13, as presented by the dashed line.  Error bars smaller than the size of symbols are not presented.} 
         \label{fig: SED}
   \end{figure*}

 {The quiescent SED (see Figure \ref{fig: SED}) has a rising slope from near to mid-infrared, consistent with an embedded young star. Differences are seen between flux measured in \textit{Spitzer} and \textit{WISE} bands, consistent with low-amplitude fluctuation seen during the quiescent stage. We estimated the bolometric luminosity of VVV-WIT-13 as 0.9 $L_\odot$, assuming a 2~kpc distance and a high line-of-sight extinction ($A_V = 17$~mag). The 0.9~$L_\odot$ luminosity is common among accreting young stars, indicating a stellar mass of 0.4 to 0.6~$M_\odot$. Here we assumed a 0.5-1~Myr of age based on the models from \citet[][]{Baraffe2015}. However, this is a rough estimation since both distance and extinction measurements have large uncertainties. When applying a closer distance (e.g. 1~kpc), the bolometric luminosity would drop by 0.60 dex, leading to a lower stellar mass estimation of 0.2~$M_\odot$.} Moreover, the absence of long-wavelength measurements ($\lambda > 24\, \mu$m) in the SED further increases the uncertainty of the luminosity estimation.

 {During the outburst, the estimated bolometric luminosity was 55~$L_\odot$, which is comparable with classical FUor-type eruptive objects (e.g. FU~Ori, V960~Mon and V733~Cep), discovered through optical surveys \citep[see Table 2 in][]{Connelley2018}. Even assuming a lower extinction (e.g. $A_V = 9$~mag), the bolometric luminosity still reaches 7.6~$L_\odot$, which remains much brighter than that of a normal low-mass pre-main-sequence star.} We note here that the outburst SED of VVV-WIT-13 has an outstanding difference compared with \textit{bona fide} FUors, as the mid-infrared outburst amplitudes of VVV-WIT-13 are comparable or even greater than the near-infrared.  However, we have so far not considered the effect of the  {variable} line-of-sight extinction. Furthermore, the geometry and evolutionary stage (i.e., protostar vs. pre-main-sequence star) are unknown, which also complicates the overall interpretation of this source.

In Figure~\ref{fig: SED}, we present the extinction-corrected SED ($\lambda F_{\lambda}$) of VVV-WIT-13. The upper limit of the extinction is set to $A_V = 17$~mag according to the extinction estimated from the near-IR colour-colour diagram (see Figure \ref{fig: CCD}). In the $A_V = 17$~mag case, we found the SED peaks in optical, with a "flat" transition between near-IR to mid-IR, which resembles the high-mass accretion scenario calculated in \citet{Bell1999}. In this model, \citet{Bell1999} assumed a hot inner accretion disk with an exceptionally high mass accretion rate at $10^{-4} M_\odot \rm yr^{-1}$, which extends to the star (see the inset in the right panel of Figure \ref{fig: SED}). In addition to the maximum at 0.6~$\mu$m, there is a secondary bump at 6~$\mu$m, due to the passive-heated disk at $\sim$0.5~au reprocessing radiation from the inner active accretion disk. In such a case, the disk temperature at 0.1~au would reach 2000~K and drop to 1000~K at a distance of 1~au. Nevertheless, there is a large uncertainty in the extinction estimation for two reasons. First, the high extinction ($A_V = 17$ mag) is estimated based on the pre-outburst near-infrared colours with empirical colours of YSOs, which might be biased by the intrinsic colours of our target. Second, there are a few examples of line-of-sight extinction that have been cleared or enhanced after an outburst of a YSO \citep[e.g.][]{Guo2024b}. We conclude that, under the assumption of $A_V = 17$ mag, the outburst SED of VVV-WIT-13 resembles a hot inner disk, which is typical for FUor-type outbursts \citep[e.g.][]{Zhu2007}. With a lower extinction estimation, such as $A_V = 9$~mag (estimated by the colour during the outburst), the SED still peaks in the mid-infrared, which does not resemble the classical FUor-type disk. In this case, we assume the existence of a warm disk, but without the presence of a hot inner part.

  \begin{figure*}
   \includegraphics[height=7.5cm]{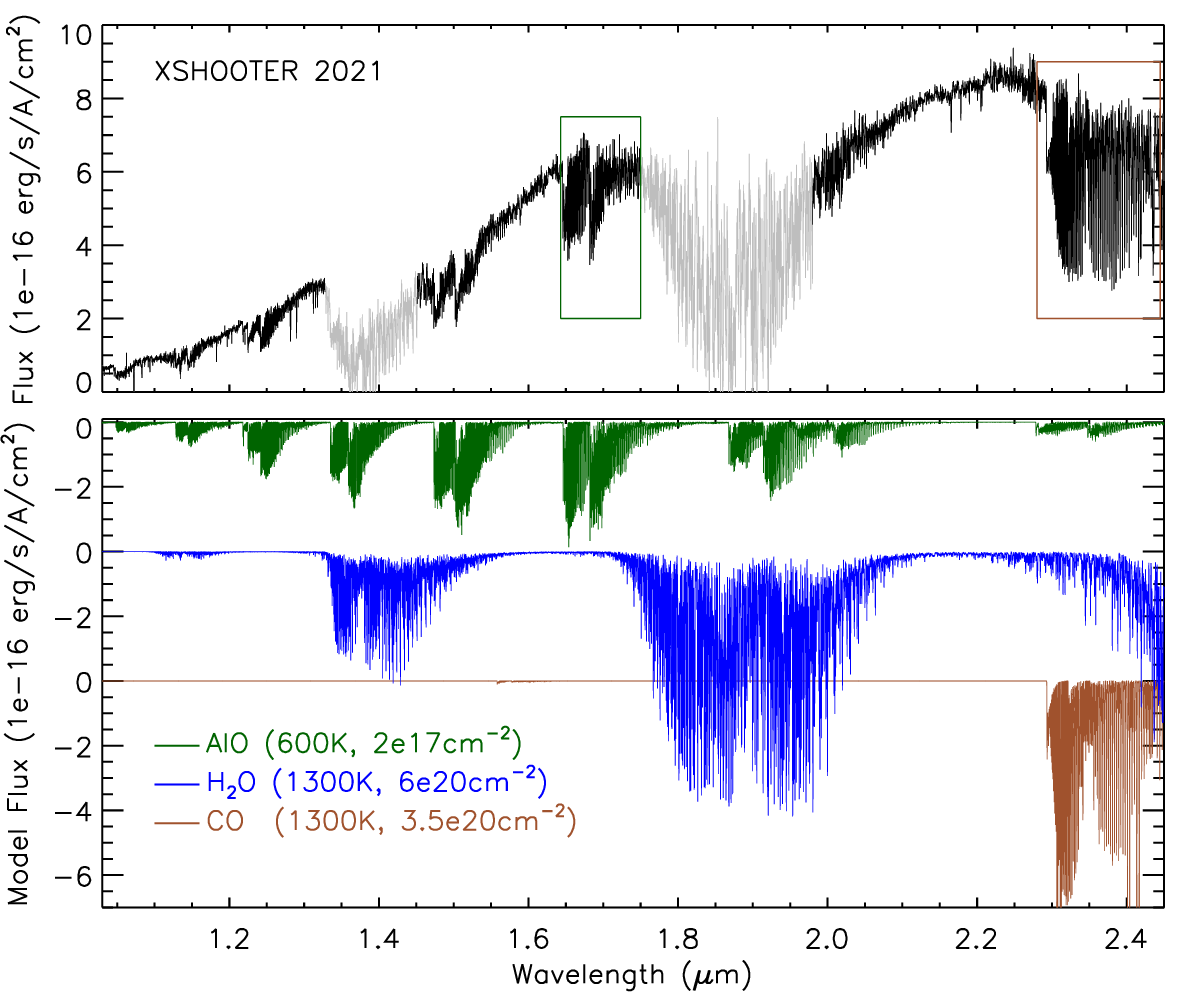}
      \includegraphics[height=7.5cm]{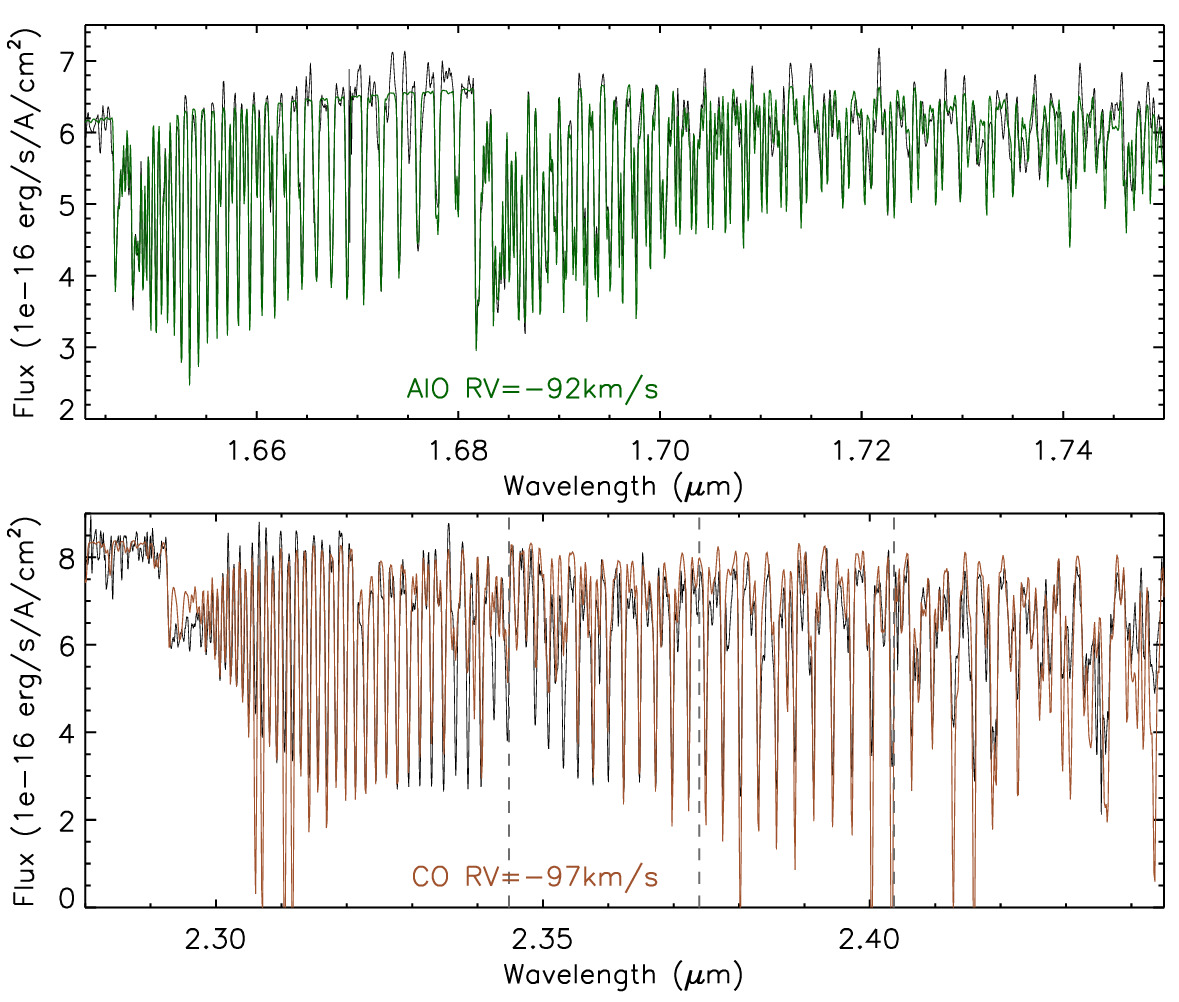}
      \caption{\textit{Upper left}: Near-infrared spectrum of VVV-WIT-13 taken during the brightness plateau. The best-fit molecular spectra (AlO, H$_2$O and CO) are presented in the \textit{lower left} panel, with effective temperature and column density labelled on the plot. The coloured squares mark the spectral regions presented on the \textit{right} panels. Spectral regions among $J$, $H$, and $K$-bandpasses are shown in grey, which are dominated by water absorption features. \textit{Right}: zoomed-in views of two regions with molecular absorption features, with radial velocity listed in the heliocentric frame. The theoretical wavelengths of $^{13}$CO absorption bandheads are presented by dashed grey lines.}
         \label{fig: spec_fit}
   \end{figure*}

\subsection{Photometric Variability}

During the pre-outburst stage, VVV-WIT-13 had low-amplitude variation with $\Delta K_s = 1.5$~mag over a 2000~d timescale. Although most YSOs exhibit photometric variation due to inhomogeneous cold/hot spots, accretion and extinction changes, such a large amplitude is uncommon. For instance, among the 442 young members of the SPICY group G342.1+0.2, only 30 sources (6.8\%) reached $\Delta K_s > 1.5$~mag (Appendix \ref{sec:AppA}). The pre-outburst variation amplitude is also greater than the majority of pre-outburst variation of long-duration eruptive YSOs \citep[see Table 4 in][]{Guo2024a}. We extracted a sinusoidal fluctuation (P = 1748$\pm141$ d, see Figure\ref{fig: spec+lc}) on the $K_s$ time series by the Lomb-Scargle periodogram \citep{Zechmeister2009}. However, the authenticity of this period can not be verified, since only one period is detected. This fluctuation is also seen in the mid-IR with $\Delta W1$= 1.6~mag and $\Delta W2$= 0.5 mag, which can not be explained as a change of line-of-sight extinction. We suspect that the pre-outburst variability is linked to the existence of a temporary disk structure as a consequence of a tidal disruption of a large body. The 1748 d period is consistent with the Keplerian rotational timescale at a distance of 2.8~au around a 1 $M_\odot$ star.

As recorded by the VVV and NEOWISE light curves, an outburst started after 25th August 2016 and rapidly reached a brightness plateau within 1 year. Limited by the cadence of the time series, the rising timescale is between 180 and 360~d. The amplitudes of the rising stage reached 3.0~mag in $W2$ and 4.5~mag in $K_s$. Such a high-amplitude and rapid-rising stage places this outburst among bona fide FUors \citep{Guo2024a}. Although the eruptive amplitude is similar to Galactic stellar merger events \citep[red novae, see][]{Kaminski2024}, the duration of the rising stage of VVV-WIT-13 is much longer than those of stellar mergers \citep[10 to 180 d, e.g.][]{Bond2003, Tylenda2011} and planet engulfment event \citep[10 d;][]{De2023}, but it is comparable to one particularly slow red nova, BLG-360 \citep{Tylenda2013}. This object is, however, very different from any other aspect than VVV-WIT-13 \citep{Steinmetz2025}.

 Approximately 1000 days after reaching its maximum, VVV-WIT-13 entered a fading stage. The duration of the outburst stage is shorter than the decades-long classical FUor-type, longer than EXor-type outbursts (timescale of 1 yr and lower infrared amplitude), and much longer than the brightness plateaus of red-novae \citep[timescale of 100 - 200 d;][]{Pastorello2019}. This intermediate eruptive timescale resembles individual events seen in V1647~Ori  \citep{Briceno2004}, VVVv322 \citep{Contreras2017}, VVV1636-4744 \citep{Guo2024a} and Source 1017 \citep{Morris2025}. Recent photometric studies reveal that this kind of intermediate-duration outburst may be as common as classical FUors and EXors among embedded sources \citep[][]{Contreras2024, Morris2025}. We infer that the accreting object of these multi-year-long outbursts must be relatively compact compared with the decades-long FUor-type events. Although we lack high-cadence photometric data, we found that the starting point of the decaying stage is not contemporaneous for the $K_s$ and \textit{WISE} bands, as the mid-IR brightness faded approximately 1~yr earlier than the near-IR. Based on the last photometric data point on Figure~\ref{fig: spec+lc}, the brightness of VVV-WIT-13 reached another plateau since 2024, about 1~mag brighter than the pre-outburst stage. Future photometric monitoring is crucial to determine the full duration of this eruptive event.

\section{Spectroscopic features and molecular models}
\label{sec:spec}

The spectroscopic features during an outburst contain crucial information such as the temperature and radial velocity of the system. The infrared spectra of most FUor-type outbursts have molecular absorption bands resembling the photosphere of cool stars \citep[][]{Connelley2018}. Specifically, CO and water vapour absorption are the most common molecular features among FUor-type outbursts. The CO gas temperature often reaches $T = 2000 - 4000 $ K \citep[e.g.][]{Contreras2017b, Liu2022}. The blue-shifted CO absorption on an eruptive YSO, V1057~Cyg, is thought to originate from an expanding shell of CO gas, ejected from the eruptive YSO at the centre \citep{Hartmann2004}. In this section, we will present detailed spectroscopic features and their evolution throughout different stages of the outburst.

\subsection{The in-outburst spectrum}

  \begin{figure*}
   \centering
   \includegraphics[height=7cm]{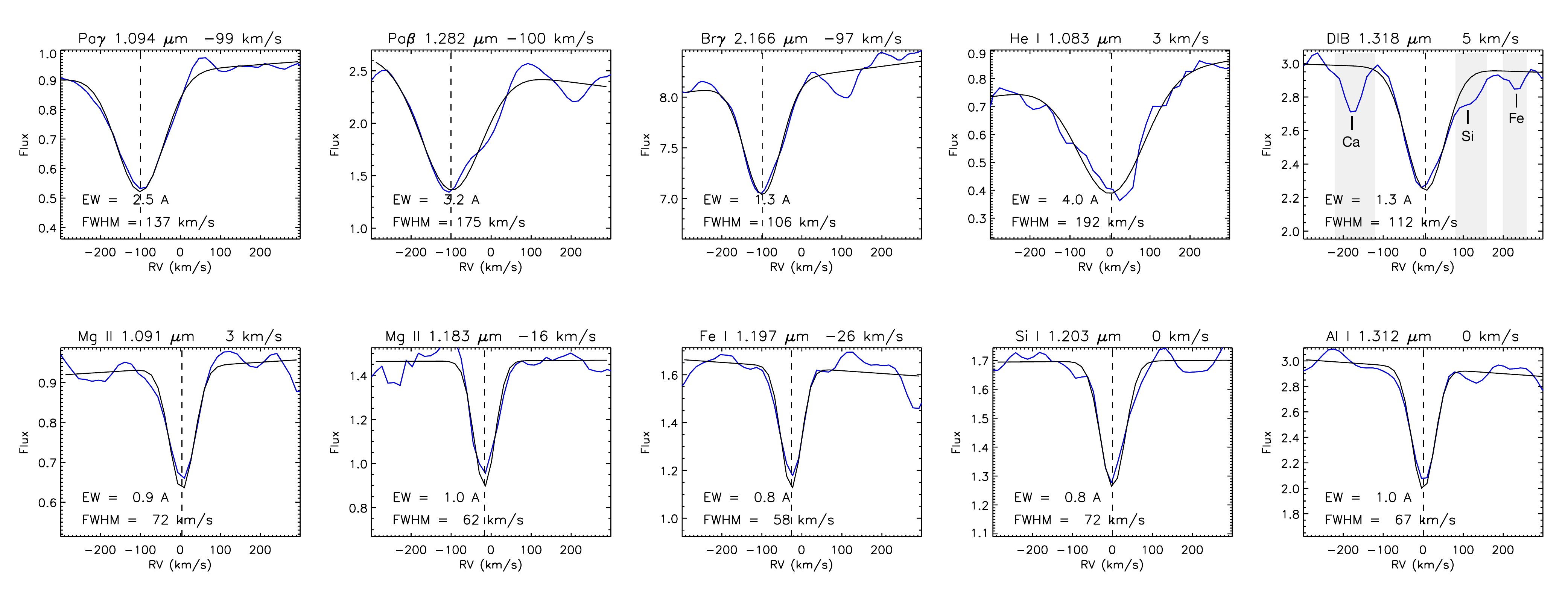}
    \caption{Absorption lines observed during the outburst stage of VVV-WIT-13. We fit each line with a Gaussian profile (shown in black), with the central radial velocity marked by dashed lines and noted on the title of each subplot (heliocentric frame). Assuming a distance of 2 kpc, the $V_{\rm LSR}$ is -22 km/s (or -15.2 km/s in the heliocentric frame). The measured EW and FWHM are labelled. In the sub-plot of DIB, the grey area is excluded from the Gaussian fitting.}
 \label{fig: Hi_spec}
   \end{figure*}
   
  \begin{figure*}
   \centering
   \includegraphics[height=4.7cm]{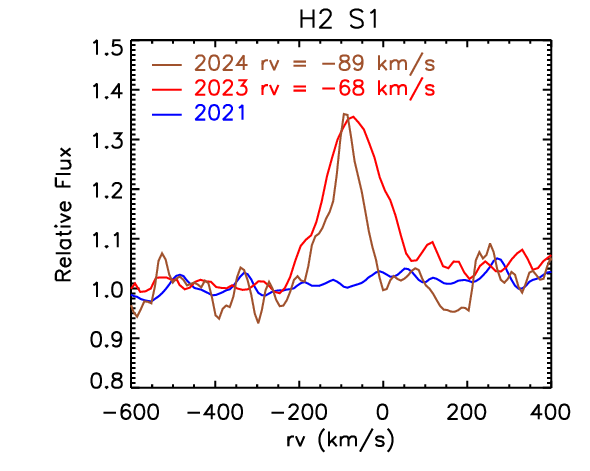}
   \includegraphics[height=4.5cm]{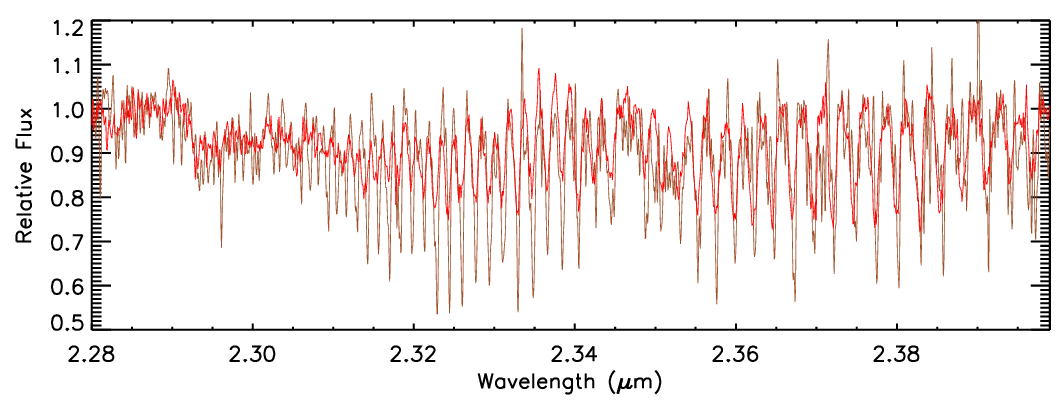}
    \caption{\textit{Left:} profiles of the 2.12 $\mu$m H$_2$ emission lines from the 2021, 2023 and 2024 epochs. \textit{Right:} K-bandpass CO-bandhead absorption features of the 2023 and 2024 spectra.}
 \label{fig: post_spec}
   \end{figure*}

During the outburst, VVV-WIT-13 displayed deep molecular absorption features, including TiO in optical, AlO in $J$ and $H$-bandpass, and $^{12}$CO bandhead beyond 2.3~$\mu$m. H~{\sc i}, He~{\sc i} and metal lines are also observed in absorption. In the following paragraphs, if not specifically noted, CO stands for the $^{12}$CO molecule and the radial velocity is measured in the heliocentric frame. Given the Galactic location of our target, the $V_{\rm LSR}$ correction from $V_{\rm helio}$ is 6.8 km/s. 

We fit the molecular absorption features using the synthetic grid spectra from the cross-sections provided by the ExoMol database \citep{Patrascu2015, Tennyson2016, Bowesman2021}. This is equivalent to absorption spectra produced by a plane-parallel slab of gas in the local thermal equilibrium and optically thin case.  We first removed the infrared continuum by applying polynomial functions to the spectrum. The continuum-removed spectrum is presented in the \textit{upper left} panel of Figure \ref{fig: spec_fit}. Four parameters were estimated for each molecular band, including the excitation temperature ($T$), radial velocity in the heliocentric frame (RV) and column density of each molecular species ($N$). The best-fit results are presented in Table~\ref{tab:spectral_features}. Compared to common FUors \citep{Guo2021}, the CO detected on VVV-WIT-13 has a low temperature ($T = 1300$~K), indicating either an envelope or outflow origin. Similarly, the AlO bands require a cool temperature of $600$~K. We will further discuss these molecular features in \S\ref{sec:molecular}. There is no obvious detection of $^{13}$CO absorption bandheads, which are usually seen among evolved stars enhanced in nucleosynthesis products \citep[see][]{Guo2024a}.  {We note that weak $^{13}$CO absorption ($\nu$'-$\nu$'', 2-0) has been observed on FU~Ori and V1057~Cyg \citep{Connelley2018}. At the 12C/13C abundance of 30-90 in the local ISM \citep{Yan2023}, a firm detection of 13CO features in the first overtone bands requires both very high sensitivity and spectral resolution, which our spectra of VVV-WIT-13 lack.}

\begin{table}[!b] 
\caption{Molecular spectral features} 
\renewcommand{\arraystretch}{1.0} 
\centering
\begin{tabular}{ccccccc} 
\hline 
\hline 
Epoch & Species & RV & T & FWHM & N\\
\hline 
  &  & km/s & K & km/s & cm$^{-2}$ \\
\hline 
2021 & CO & -97 & 1300 & - & 3.5e20 \\
2021 & AlO & -92 & 600 & - & 2.0e17 \\
2023 & CO & -98 & 1500 & - & 0.9e20 \\
2023 & $H_2$ & -89 & - & 156 &\\
2024 & $H_2$ & -68 & -  & 88 &\\
\hline 
\hline 
\end{tabular}
\tablefoot{The expected heliocentric system velocity is -15 km/s.}

\label{tab:spectral_features}
\end{table}

We examined H~{\sc i} absorption lines by fitting Gaussian profiles (Pa$\,\beta$, Pa$\,\gamma$ and Br$\,\gamma$, see line profiles and properties on Figure \ref{fig: Hi_spec}). All lines share similar RV with AlO and CO, ranging from -94 to -100~km/s, and equivalent widths (EW) are 3.4, 2.6 and 1.1 \AA$\,$, respectively.  The Gaussian FWHM of the line profiles is broader than the Gaussian broadening of molecular bands and the spectral resolution (15 km/s). The Pa$\,\beta$ absorption line has an EW of 3.4 \AA, placing it at the higher end of values from FUors \citep[see][]{Connelley2018}. If we assume a distance of 2.0$\pm$1.0 kpc, the kinematic velocity at the Galactic longitude of VVV-WIT-13 is between -37~km/s and -10~km/s \citep[in LSR;][]{Wenger2018}. Hence, the molecular bands and  H~{\sc i} lines are blue-shifted.

 {The broad He~{\sc i} absorption line is detected at 1.083~$\mu$m, as another pronounced feature among FUors, as a tracer of wind. However, the He~{\sc i} line does not have a significantly blue-shifted central radial velocity (3~km/s), which is inconsistent with other bona fide FUors. We also detected a group of metal lines, including Mg~{\sc ii}, Fe~{\sc i}, Si~{\sc i} and Al~{\sc i}. All of these metal lines have symmetric and narrow line profiles, with low RV, consistent with photospheric lines detected among FUors. }

\subsection{The fading spectrum}

 {The two spectra in the 2023/2024 epochs have lower continuum flux than the 2021 epoch, as the $K_s$ brightness was 3~mag fainter (Figure \ref{fig: post_spec}). We found that the CO absorption bands in 2023/2024 epochs are shallower than the 2021 epoch, with a lower column density (see Table \ref{tab:spectral_features}). No AlO absorption is detected, indicating that the deep AlO absorption bands are temporary features associated with the eruptive event. }

We detect a blue-shifted H$_2$ $1-0$ S(1) emission line at 2.12~$\mu$m, which is a common indicator of stellar wind or outflow in accreting YSOs \citep[][]{Greene1996, Greene2010}.  {This line is from an upper energy level 6956 K above the ground and, in principle, requires much higher excitation than CO or AlO observed in absorption. }Such an emission line feature is not commonly seen among FUor-type eruptive YSOs. Instead, it has been observed among EXors \citep[e.g.][]{Park2022} and a few peculiar eruptive YSOs \citep[e.g.][]{Kospal2020}. The combination of H$_2$ emission and CO bandhead absorption features is only seen in a few cases, including the periodic variable EC 53, FUor-candidates without observed outbursts \citep[e.g. CB 230 and IRAS 18270–0153W;][]{Connelley2018}, and a rapidly faded eruptive YSO VVVv322  \citep[$\Delta K_s = 2.6$ mag;][]{Contreras2017b, Contreras2017}. We fit simple Gaussian functions to the H$_2$ line profiles of the 2023 and 2024 epochs, with measured RVs as -89~km/s and -68~km/s, FWHM as 156~km/s and 88~km/s, and equivalent widths as -53 \AA$\,$ and -30 \AA. 
Such broad and variable H$_2$ lines were previously observed on a few variable YSOs \citep[e.g. VVVv815;][]{Guo2020}, tracing temporary features in the stellar wind/outflow. The RV of H$_2$ lines are redder than molecular bands, which is uncommon on eruptive YSOs.  {Alternatively, it has been observed on a red nova \citep{Steinmetz2025}.}   {Because absorption lines trace only material along the line of sight to the continuum source, the variations in the RV measured from molecular bands of the oxides trace small changes in the velocity field where these bands are excited. The H$_2$ emission presumably arises from a larger volume of shocked gas and thus is expected to be less blueshifted if the overall velocity field is dominated by expansion.
} 

In summary, VVV-WIT-13 exhibited unique spectral features in both the outburst and fading stages. During the outburst stage, it displayed deep absorption features including water vapour, AlO and CO molecular bands, resembling a luminous cool accretion disk. H~{\sc i} lines are detected in absorption during the outburst stage, with similar RV to molecular bands. After the outburst, with the fading/cooling of the continuum emission, the AlO bands are no longer detected in the spectra, whilst the CO bandhead absorption still exists with a similar RV to the outburst stage. We observed broad and variable H$_2$ emission associated with stellar wind. VVV-WIT-13 is the first high-amplitude eruptive source exhibiting H$_2$ emission and CO absorption simultaneously. The 2023 and 2024 spectral features indicate a transitional phase between the eruptive and the quiescent stages. 

\label{sec:molecular}

\section{Discussion}
\label{sec:discussion2}  

\begin{figure*}
   \includegraphics[height=7.5cm]{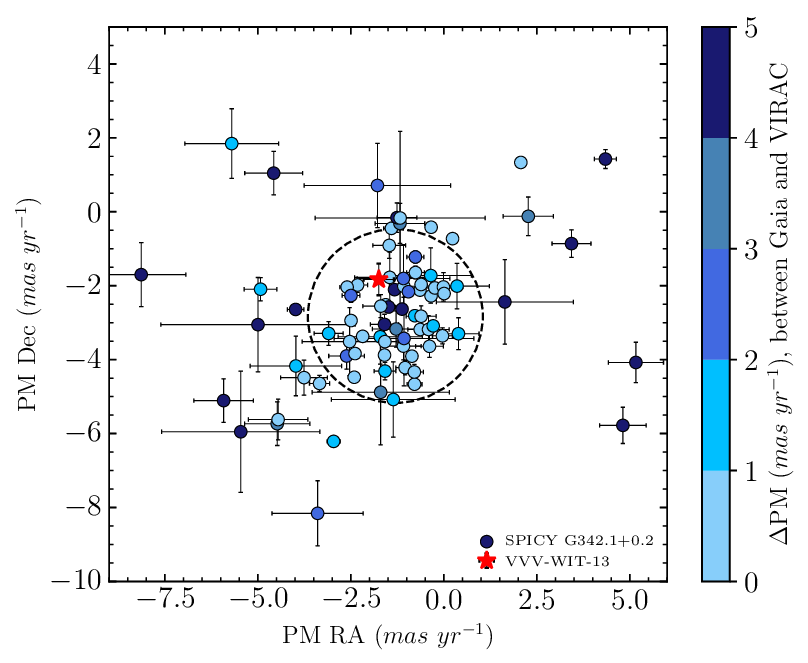}
      \includegraphics[height=7.5cm]{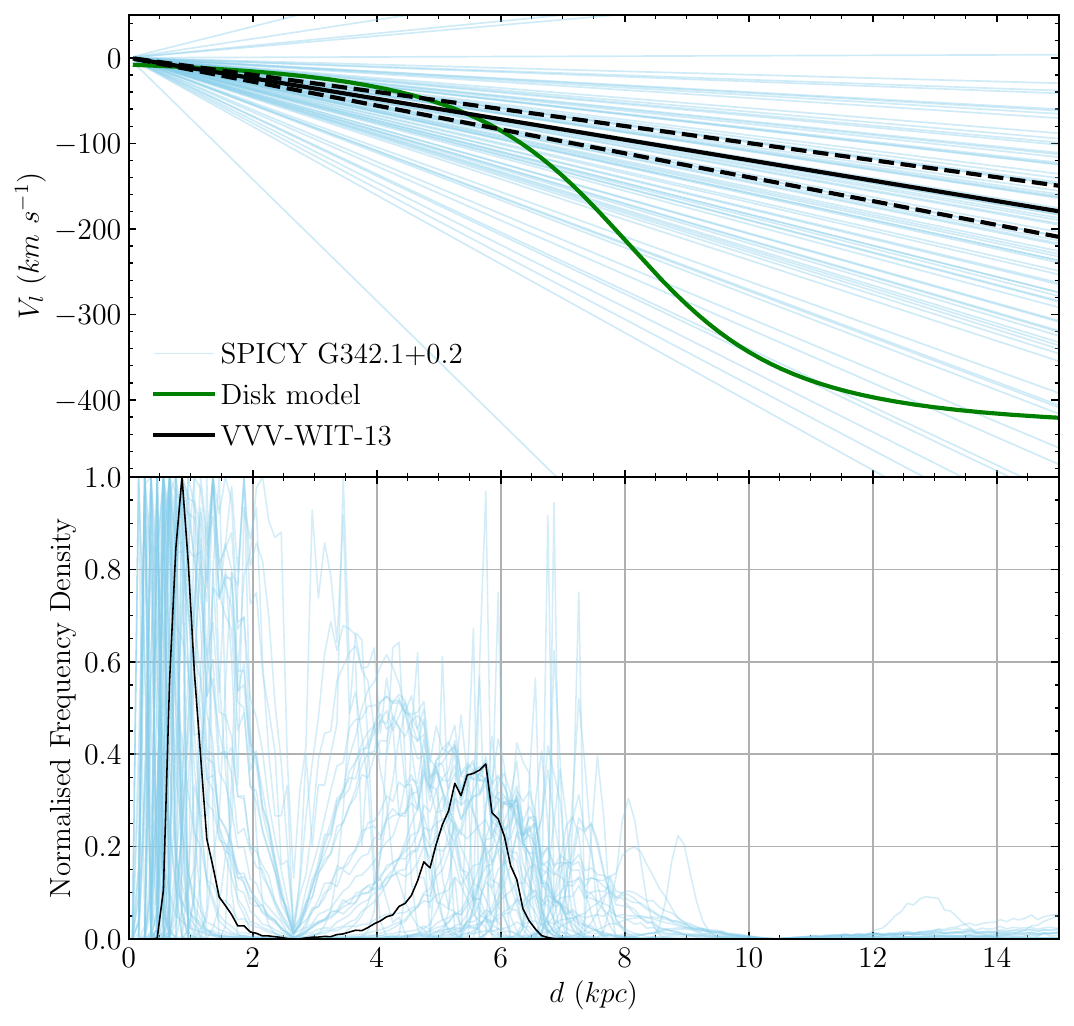}
      \caption{\textit{Left}: the \textit{Gaia} proper motions (PMs) of 83 group members of SPICY G342.1+0.2. The dashed ring marks the 1-$\sigma$ distribution of PMs from the centre of the group. The PMs of VVV-WIT-13 are obtained from the VIRAC2 catalogue of the VVV survey. The colour bar represents the difference between the PMs measured from \textit{Gaia} and VVV. \textit{Right}: the PM distance of VVV-WIT-13 (black) and the 83 sources presented on the \textit{left} panel.}
   \label{fig: PM_gaia}
   \end{figure*}

\subsection{The outflow origin of molecular bands}

Absorption and emission molecular bands of metal oxides are relatively rare in eruptive YSOs. Some notable examples are TiO and VO bands in emission in a few embedded Class I YSOs \citep{Hillenbrand2012, Hillenbrand2013}. While TiO and VO are common in late-type stellar atmospheres, photospheric and circumstellar AlO are rare. The gas phase of AlO has been detected in O-rich circumstellar envelopes, such as VY~CMa \citep{Tenenbaum2009, Kaminski2013} and Miras \citep[reviewed by][]{Kaminski2019}. However, AlO is highly refractory, as it can quickly condense into solid dust grains, such as Al$_2$O$_3$. In chemical equilibrium at densities of a red supergiant wind or around a bloated star with a high mass accretion rate, the AlO molecule only forms efficiently at temperatures of 1100 K (with $n_H \sim 10^4 \rm cm^{-3}$). The presence of AlO indicates that it must have been preserved in the gas phase within a wind/shock environment, allowing it to survive below its expected formation temperature. Therefore, the chemical chain was disrupted by the rapid expansion of the gas, leading to the survival of the AlO molecule \citep{McCabe1979, Tenenbaum2009}. Alternatively, sputtering in shocks or sublimation may also release AlO to the gas phase from solids. 

As stated in \S\ref{sec:spec}, AlO bands were observed in the ejecta of stellar mergers, but deep AlO absorption has not been reported on young stars in the near-infrared, which raised our concern about the nature of VVV-WIT-13.  In the sub-mm wavelength, the AlO emission was specially resolved by ALMA around a high-mass young star, revealing that AlO is present at the base of the outflow \citep{Tachibana2019}. Here, we consider that both CO and AlO absorption bands arise from an expanding shell {/wind} from VVV-WIT-13. This is similar to what has been observed on V1057 Cyg, in which case the blue-shifted CO absorption bands are thought to originate from the ejecta of the eruptive YSO at the centre \citep{Hartmann2004}. It is also similar to the shell-expanding scene observed among red novae.

This outflow-formation scenario would explain the blueshifted radial velocity of CO and AlO molecules against the Galactic rotational velocity and the metal absorption lines (see Figure~\ref{fig: Hi_spec}). The strong outflow is commonly detected among eruptive variables, including FUors and red novae, which is traced by blue-shifted absorption lines (e.g. He~{\sc i}).  {Here, we estimated the mass loss rate of VVV-WIT-13 based on the column density and velocity of CO measured in this work ($N_{\rm CO} = 3.5\times10^{20} \rm cm^{-2}$) and ($V_{\rm CO} \sim -100\, \rm km/s$). We also assume that the wind is composed of 70\% of hydrogen and 30\% of helium (average atomic mass is 1.3), with a typical CO abundance ($X_{\rm CO} = n_{\rm CO}/n_{\rm H_2} = 10^{-4}$). Therefore, the mass lost rate can be written as
\begin{equation}
    \dot{M} \;\simeq\; 4\pi\, r_{\rm CO}\, \mu m_{\rm H}\, \frac{N_{\rm CO}}{X_{\rm CO}}\, V_{\rm CO}\, \eta\,,
\end{equation}
where $\mu = 1.3$ and $m_{\rm H} = 1.67\times10^{-27}$~kg. The covering fraction of the wind against a full solid angle is parameterised by $\eta$, which has a common value of 0.2 assuming an open angle of 40 deg. We assumed the CO wind is launched at $r_{\rm CO}$ = 0.07~au from the star, by matching the escape velocity of 100~km/s around a 0.4 $M_\odot$ young star. The mass loss rate is on the order of $10^{-5}\,\ M_\odot\rm{yr^{-1}}$, comparable to the mass loss rate observed on FU~Ori \citep{Calvet1993}. We note the mass accretion rate on FUors is normally on the order of $10^{-5}$ to $10^{-4}\,\ M_\odot\rm{yr^{-1}}$. This mass loss rate is also comparable to the post-main-sequence AGB stars, which is between $10^{-6}$ and $10^{-4}\,\ M_\odot\rm{yr^{-1}}$ \citep{Hofner2018}. However, it is orders of magnitude smaller than the mass loss rate during the red novae stage \citep{Tylenda2013, Blagorodnova2021}.}

\subsection{The extinction measurements through spectral features}
\label{sec:dib}
Diffuse Interstellar Bands (DIBs) are broad absorption bands, mainly contributed by organic compounds in the interstellar clouds \citep[see e.g.,][]{Campbell2015}. We found a DIB on the 2021 epoch spectrum of VVV-WIT-13 at 1.318~$\mu$m, but possibly blending with a nearby Si {\sc{i}} line \citep{Origlia2019}. We fit the DIB with a Gaussian profile across wavelength ranges that avoid including contributions from metal lines (see the top right panel of Figure \ref{fig: Hi_spec}). We measured an EW of 1.3 {\AA} and an RV of 5~km/s for this DIB. The RV of the $\lambda$1.3178~$\mu$m DIB absorption is redder than the H {\sc i} lines, and molecular absorption bands and comparable to photospheric metal lines. Based on the empirical correlation between the DIB EW and the dust reddening derived by \citet{Hamano2016}, we estimated $E(B-V) \approx 2.9$ mag, corresponding to an $A_V \sim 9.0$~mag assuming $R_V=3.1$. The choice of $R_V$ is an average of the Galactic diffuse interstellar medium, consistent with the extinction law we used for this work.  {This likely represents only the interstellar reddening component on the line of sight and does not include the local extinction from any circumstellar material surrounding VVV-WIT-13.}

The ratio between the fluxes of two H$_2$ lines is another extinction indicator as the theoretical value is independent of the excitation conditions \citep[$1-0$~Q(3)/S(1)~=~0.7;][]{Turner1977}. According to the extinction curve from \citet{WangS2019}, we have $A_V = 130\log_{10}(f_{H_2}/0.7)$, where $f_{H_2}$ is the measured flux ratio of two H$_2$ lines. In our 2023 and 2024 spectra, due to the low signal-to-noise ratio beyond 2.42~$\mu$m (S/N = 5 and 2, respectively), we can only roughly estimate the extinction through this method. Using the spectrum taken in 2023, we measured $f_{H_2} = 0.95 \pm 0.15$, and therefore $A_V = 17.2 \pm 8.3 $~mag. Although with a large uncertainty, this measured $A_V$ is consistent with the extinction measured on the colour-colour diagram during the quiescent stage, and it is larger than the 9~mag extinction estimated from DIB observed during the outburst. Here, we can not completely rule out the possibility that the circumstellar extinction has increased during the fading stage, due to the circumstellar dust pile-up again, whilst the disk is cooling down. We also suspect that the extinction measured from DIB is the interstellar extinction instead of the circumstellar extinction. 

\subsection{The Galactic location of VVV-WIT-13}
\label{sec:discussion_group}

 Here, we investigate the nature of VVV-WIT-13 by examining its Galactic location to obtain a more accurate distance modulus. We first test the hypothesis that VVV-WIT-13 shares the same distance as members from the SPICY YSO group G342.1+0.2. In \citet{Kuhn2021}, groups of YSOs were identified through spatial clustering of YSOs using the HDBSCAN algorithm, which was originally applied to detect Galactic open clusters in Gaia DR2 \citep[e.g.][]{Kounkel2019}. However, the classification relied solely on 2D Galactic coordinates and did not incorporate distance or kinematic information. To address this limitation, we examine the proper motions (PMs) of both the SPICY group and VVV-WIT-13 using data from Gaia DR3 and the VVV survey's VIRAC catalogue \citep[the VIRAC2 version;][]{Smith2025}.

 We cross-matched the coordinates of the G342.1+0.2 group members with the Gaia DR3 using a matching radius of 1\arcsec. A total of 83 SPICY group members have Gaia PM measurements. We then retrieved the VVV PMs of these sources from VIRAC (see Appendix). For most sources, Gaia and VVV PMs differ by less than 2 mas/yr, with larger discrepancies occurring only for those with high uncertainties of the PM measurements (Figure~\ref{fig: PM_gaia}). VVV-WIT-13 lies only $\sim$1 mas/yr from the projected centre of the SPICY group, suggesting a kinematic link to the group members. The $\sim$1 mas/yr threshold is commonly used for identifying open cluster members at 2 kpc \citep[e.g.][]{Borissova2018}. Using VIRAC PMs, we estimated a PM-based distance of VVV-WIT-13 following the method of \citet{Guo2021}, finding two likely ranges: 0.3–2 kpc or beyond 4 kpc (see the right panel of Figure~\ref{fig: PM_gaia}). The close distance range is more reliable, as the Galactic rotational velocity is -80 km/s at 4-5 kpc, which is much larger than the RV observed on the photospheric lines and cannot be explained by non-circular motions.

 Given the complexity of establishing spatial and kinematic associations along the Galactic plane, we then assessed whether SPICY group G342.1+0.2 is truly dynamically bound. We searched Gaia DR3 sources located within $15\arcmin$ of VVV-WIT-13 with $G < 19$ mag (5961 sources). The $G$-band brightness limit is set to eliminate sources with large PM and parallax ($\varpi$) uncertainties. We found that 37\% of Gaia sources have PMs concentrated within the 1-$\sigma$ region shown in Figure~\ref{fig: PM_gaia}. Among them, more than 1700 sources have $\varpi > 0.33$ mas/yr, corresponding to distances $<$3 kpc. The number of Gaia sources sharing the same spatial and kinematic measurements as the SPICY group is an order of magnitude larger than the SPICY group itself. Therefore, we conclude that SPICY G342.1+0.2 is a loose YSO association rather than a bound cluster. Nevertheless, this does not affect our subsequent investigation of VVV-WIT-13’s Galactic location.

  \begin{figure}
   \centering
   \includegraphics[width=8cm]{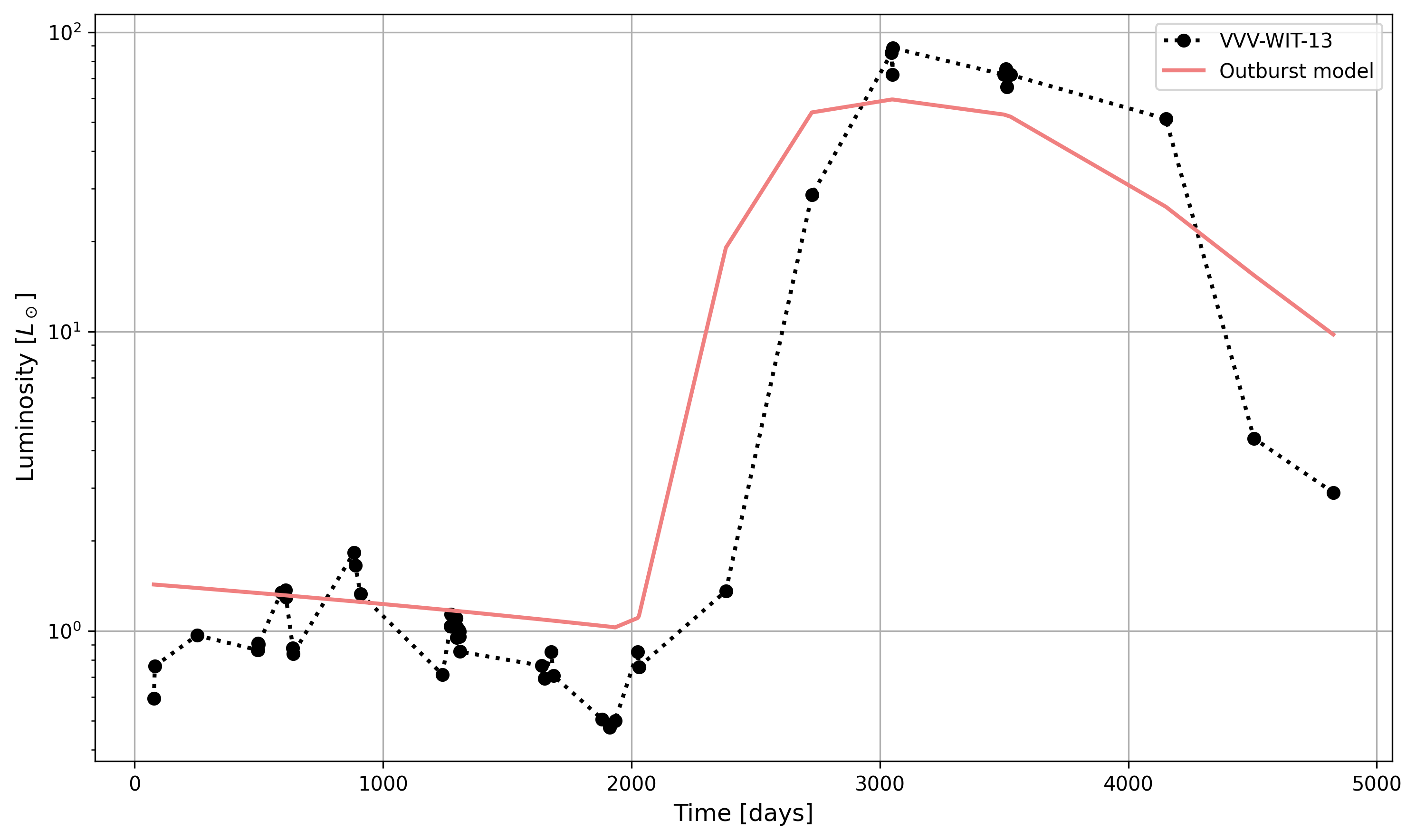}
      \caption{Time evolution of the bolometric luminosity of the system. The black curve (dotted line) shows the observational data from VVV-WIT-13 (converted to bolometric), while the red curve (solid line) corresponds to our hydrodynamic model. The time axis of the model has been shifted to align the luminosity peaks. The agreement between the rise and decay timescales suggests that the observed outburst is consistent with the tidal disruption of a giant $5 M_J$ mass planet embryo.} 
         \label{fig: luminosity}
   \end{figure}
   
 At $l = 342^\circ$, the Galactic spiral arm model places the Sagittarius and Scutum–Centaurus arms at distances of 1 kpc and 3 kpc, respectively \citep{Reid2019}. Most retrieved Gaia sources, including SPICY YSOs, are thus likely located within or nearer than the Scutum–Centaurus arm, consistent with Gaia’s limited sensitivity to distant, embedded and faint sources ($G > 19$ mag). Placing VVV-WIT-13 further beyond (e.g. in the Norma arm or Galactic bulge) would imply interstellar extinction $A_V > 10$ mag \citep[see extinction maps;][]{Soto2019, Zucker2025}\footnote{This estimation of A$_V$ does not include the extinction caused by the infrared dark cloud 342.135+0.204, due to the width of the cloud is narrower than the resolution of the extinction map.}. The distance modulus at a 5 kpc distance is 13.5 mag, pushing VVV-WIT-13 into the intermediate-mass YSO regime (M$_K <$ 1 mag and mass $> 1.5$ $M_\odot$), where such accretion bursts are extremely rare. We also note that VVV-WIT-13 is projected against a Galactic infrared dark cloud. Such clouds are the densest regions of molecular clouds, and the probability of a chance alignment in the VVV Galactic disk fields is only $\sim$1\% \citep[Section 4 in][]{Lucas2020WIT}. In summary, the kinematic and location information support the classification of VVV-WIT-13 as a Galactic young star with a distance less than 3 kpc.

 Nevertheless, we explore the possibility that VVV-WIT-13 is a red nova. The progenitors of red novae are usually contact binaries or systems entering common-envelope evolution \citep[e.g. V1309 Sco][]{Kaminski2015}. Those sources are not expected to have a significant mid-infrared excess, which does not fit the pre-outburst SED of VVV-WIT-13. Alternatively, some luminous red novae have progenitors that appear to be evolved massive stars, which could have previous ejecta or a dusty envelope corresponding to the mid-infrared excess \citep[see examples in][]{Blagorodnova2021}. VVV-WIT-13 is clearly too faint to be a high-luminous event (e.g.  $10^4\, L_\odot$), unless located on the other side of the Galaxy (10 kpc) with a high extinction (e.g. A$_V >$ 30 mag). In such a case, the proper motion of this source will be significantly different from the observed values (see the right panel of \ref{fig: PM_gaia}). However, VVV-WIT-13 can still be a low-luminosity red nova that happened on an embedded post-main-sequence star.

 An interesting relation to red novae and luminous red novae is that VVV-WIT-13 displayed a short brightening event before the main eruption. The brightening event occurred at around day $\Delta t$=1000, i.e. some 2000 days before the beginning of the main eruption (see Figure \ref{fig: spec+lc}). In some red novae, this early short-lasting rise in brightness is also observed before the main outburst and is often referred to as the "precursor" \citep{Addison2022}. Its origin is not yet known, but it has been associated with the L2 mass loss and dynamical interaction between the two bodies, which eventually merge, producing the main red nova outburst \citep[e.g.][]{Pejcha2017, Metzger2017} Although this is not a standard scenario now considered for red novae, some authors ascribe this phase of the merger to the tidal disruption of the secondary (e.g. a planet) whose shattered material is later accreted onto the primary, producing the main burst \citep{OConnor2023, Yarza2025}.

\subsection{A tidally disrupted planet embryo?}
\label{sec:discussion}

 {In this work, we found some clues suggesting that the outburst on VVV-WIT-13 might be a disruption of a giant planet embryo around a young star. First, assuming Keplerian rotation around a 1 $M_\odot$ star, the 1748 d pre-outburst variation timescale corresponds to a $r_{\rm KEP} = 2.8$~au orbit (or 1.8~au around a 0.5~$M_\odot$ star), which may be interpreted as an asymmetric warm accretion disk. Second, it explains the high mid-infrared amplitude, as the blackbody emission from a warm disk ($T=850$~K) peaks at 3.4~$\mu$m according to Wien's displacement law. Third, the duration of the outburst is 2000 to 3000 days, comparable to the dynamic timescale at such a radius. This timescale is much longer than the timescale of EXor-type outbursts caused by instability at the innermost disk and the classical red nova ($200 - 300$ days). In this scenario, the warm disk ring does not reach the stellar surface, with an inner cavity. Therefore, it maintains an intermediate effective temperature, which explains the temperature estimated from molecular bands (see \S\ref{sec:spec}). This scenario predicts that the initial outburst occurs due to the dynamical rearrangement of the orbits of the disrupted material of a low-density planet embryo. During this time, the innermost disk regions do not flare as they are spatially distinct from the region where the disruption occurs. This flare thus has a very different disk structure: a hot ring in the disk versus a hot inner disk in FUors.
}

 To test this theoretical scenario and compare with observations, we performed 2D hydrodynamic simulations using FARGO3D \citep{FARGO3D2016}. Additionally, our model includes a time-dependent energy equation under the adiabatic approximation. This allows us to compute the effective temperature of the disk and subsequently estimate the radiative flux from the surface. In summary, our ad-hoc hydrodynamic model reproduces not only the amplitude of the outburst but also the rise/decay timescales and the duration of the eruptive event (see Figure \ref{fig: luminosity}
). Although our current attempts are only based on ad-hoc models, our results indicate that a tidally disrupted $5 M_J$ mass planet embryo is a possible explanation for the outburst observed on VVV-WIT-13. If correct, this would represent the first direct observational evidence of such a phenomenon, offering a unique window into the early dynamical evolution of massive clumps in protoplanetary disks \citep{Nayakshin2010a, Cha2011}. However, we acknowledge some drawbacks of this model. For example, the formation and migration history of the embryo is not thoroughly discussed, which therefore affects the plausibility of this scenario. The detailed methods of the simulation, more sophisticated exploration of the parameter space, and careful calculations on radiative transfer will be presented in a separate publication (Montesinos et al., in prep). 

\section{Summary}
\label{sec:summary}
In this work, we present photometric and spectroscopic analyses of an infrared eruptive source: VVV-WIT-13.  {We estimated the distance of VVV-WIT-13 as 2 kpc, considering it is a member of the SPICY YSO group G342.1+0.2. Initially, this target was classified as an eruptive YSO based on its photometric characteristics and pre-outburst infrared SED studies \citep[SPICY][]{Kuhn2021}. 
An outburst is observed on VVV-WIT-13 since the second half of the year 2016, with amplitudes $\Delta K_s = 4.5$ mag, $\Delta W1 = 3.9$~mag and $\Delta W2 = 3.0$~mag. The rising stage of the outburst is only captured by 1-2 epochs in the infrared photometry, with a duration of less than 1 year. }

Three sets of spectra were observed during the event, with many absorption features, including H~{\sc i}, CO, H$_2$O and AlO absorption bands. The CO and AlO bands have RV around -100 km/s, significantly bluer than photospheric lines (e.g. Si{\sc i} and Al~{\sc i} at 0~km/s). We suspect that the temporary AlO bands observed during the outburst were formed within the stellar wind/outflow, and thus the AlO molecules are preserved from condensing into the dust form. The AlO absorption bands have never been recorded on an outbursting YSO, and they were previously seen among red novae. The true nature of VVV-WIT-13 is still unclear, and hence the name "What Is This?".

We found the following evidence that indicates VVV-WIT-13 is an outbursting young star:
\begin{itemize}
    \item  {VVV-WIT-13 is projected in a Galactic star-forming region with several Galactic molecular clouds, \textit{Herschel} Hi-Gal clumps and H {\sc ii} regions within 5$\arcmin$ radius. It is located within or behind an infrared dark cloud, G342.135+0.204.}

    \item  {We studied the spatial and dynamical association of VVV-WIT-13 and discovered that the VVV VIRAC proper motions of VVV-WIT-13 are consistent with a distance between 0.3 to 2 kpc. The radial velocity of VVV-WIT-13, measured from photospheric lines, supports this distance estimation rather than a farther Galactic location.}

    \item  {During the quiescent stage, VVV-WIT-13 had infrared colours consistent with an embedded YSO. Using the empirical colours of YSOs, we estimated a pre-outburst line-of-sight extinction $A_V = 17$ mag, and a bolometric luminosity of 0.9 $L_\odot$. According to pre-main-sequence evolutionary models, VVV-WIT-13 is a low-mass YSO ($M = 0.4 - 0.6\, M_\odot$). Some low-amplitude variations are observed on the pre-outburst light curves, with a possible 1748-day period, which might be linked to an asymmetric circumstellar disk.}

    \item  {The duration of the outburst observed on VVV-WIT-13 is $\sim$2000 days, which is intermediate between the year-long EXor-type outbursts and the decades-long FUor-type outbursts. This places VVV-WIT-13 among the new category of infrared observed multi-year-long outbursts \citep[reviewed by][]{Fischer2023}. The full-event timescale of VVV-WIT-13 significantly exceeds that of most luminous red novae, which are typically shorter than 1000 days.} 

    \item  {Using the CO bandhead absorption feature, we estimated the mass loss rate during the outburst is on the order of
    $\rm 10^{-5}M_\odot/yr$, which is comparable to FU Ori but much smaller than stellar merger events.}
\end{itemize}

The fading stage of VVV-WIT-13 started after 2021, and the $K_s$-band brightness rapidly dropped below 14~mag in two years. Two spectra taken during the fading stage no longer have AlO absorption features, but with broad H$_2$ emission lines. The CO bandhead absorption feature is also shallower than during the outburst stage. 
     
 We proposed one possible theoretical solution to the outburst observed on VVV-WIT-13 as a tidally disrupted planetary embryo. Our ad-hoc model can recreate the observed outburst. A forthcoming paper will explore the parameter space and detailed theoretical models. We note that there are still many unknown facts about this source, and the disruption of a planetary embryo is only one theoretical solution.

 In a recent spectroscopic observation, the CO bandheads beyond 2.3~$\mu$m turned from absorption into emission features, suggesting the existence of hot CO gas near the central star. Common indicators of magnetospheric accretion, such as Br$\gamma$ and Na {\sc i} doublets, are detected, which further confirms the young nature of this source. The mystery about VVV-WIT-13 continues, and the most recent inner disk evolution and theoretical simulations will be presented in a forthcoming publication. In addition, observations at long wavelengths would help to define other important parameters of this source, such as disk inclination angle, the disk mass and the existence of sub-structures in the circumstellar disk.

\begin{acknowledgements}
We thank the constructive comments from the anonymous referee. We gratefully thank the helpful comments from Prof. Gregory Herczeg. This work is supported by the China-Chile Joint Research Fund (CCJRF No.2301) and the Chinese Academy of Sciences South America Center for Astronomy (CASSACA) Key Research Project E52H540301. CCJRF is provided by the CASSACA and established by the National Astronomical Observatories, Chinese Academy of Sciences (NAOC) and Chilean Astronomy Society (SOCHIAS) to support China-Chile collaborations in astronomy. ZG, RK, JB and JO are funded by ANID, Millennium Science Initiative, AIM23-001. ZG and CM are funded by the project ALMA-ANID 31240014. MM acknowledges financial support from FONDECYT Regular 1241818. JB and RK thank the support from FONDECYT Regular project No. 1240249. R.K.S. acknowledges support from CNPq/Brazil through projects 308298/2022-5 and 421034/2023-8. A.C.G. acknowledges support from PRIN-MUR 2022 20228JPA3A “The path to star and planet formation in the JWST era (PATH)” funded by NextGeneration EU and by INAF-GoG 2022 “NIR-dark Accretion Outbursts in Massive Young stellar objects (NAOMY)” and Large Grant INAF 2022 “YSOs Outflows, Disks and Accretion: towards a global framework for the evolution of planet forming systems (YODA)”.

VE acknowledges support from the Ministry of Science and Higher Education of the Russian Federation (State assignment in the field of scientific activity 2023, GZ0110/23-10-IF).

LZW and LXY  are supported by the Chinese Academy of Sciences South America Center for Astronomy (CASSACA) Key Research Project E52H540301, and in part by the Chinese Academy of Sciences (CAS) through a grant to the CASSACA.
C.C.P. was supported by the National Research Foundation of Korea (NRF) grant funded by the Korean government (MEST; No. 2019R1A6A1A10073437)"

This publication makes use of data products from the Near-Earth Object Wide-field Infrared Survey Explorer (NEOWISE), which is a joint project of the Jet Propulsion Laboratory/California Institute of Technology and the University of Arizona. NEOWISE is funded by the National Aeronautics and Space Administration. This research has made use of the NASA/IPAC Infrared Science Archive, which is funded by the National Aeronautics and Space Administration and operated by the California Institute of Technology.

This work has made use of the University of Hertfordshire's high-performance computing facility (\url{http://uhhpc.herts.ac.uk}).

This research has made use of the SVO Filter Profile Service "Carlos Rodrigo", funded by MCIN/AEI/10.13039/501100011033/ through grant PID2020-112949GB-I00.

The authors acknowledge the use of artificial intelligence language models (ChatGPT, Deepseek and Grammarly) for assistance with text editing.
\end{acknowledgements}

\bibliographystyle{aa}
\bibliography{reference}
\begin{appendix}

\section{Photometric measurements}
Here, we present a table containing photometric measurements of VVV-WIT-13. A full version of the table (machine-readable format) is attached as a
supplementary file. 
\begin{table}[!b] 
\caption{Photometric data} 
\renewcommand{\arraystretch}{1.31} 
\centering
\begin{tabular}{ccccccc} 
\hline 
\hline 
Time & Filter & Origin & Brightness & Error & \\
\hline 
MJD  &  & & mag & mag & \\
\hline 
\multicolumn{2}{l}{Pre-outburst}\\
\hline 
51321 & $K_s$ & 2MASS & 14.95 & 0.55 & p\\
53253 & $IRAC1$ & GLIMPSE & 11.96 & 0.05 & c\\
53253 & $IRAC2$ & GLIMPSE & 10.75 & 0.06 & c\\
53253 & $IRAC3$ & GLIMPSE & 9.51 & 0.05 & c\\
53253 & $IRAC4$ & GLIMPSE & 8.58 & 0.05 & c\\
54031 & [24] & MIPS24 & 5.25 & 0.10 & c\\
55262 & $W1$ & ALLWISE & 13.64 & 0.24 & c\\
55262 & $W2$ & ALLWISE & 10.89 & 0.04 & c\\
55262 & $W3$ & ALLWISE & 8.74 & 0.20 & c\\
55262 & $W4$ & ALLWISE & 6.08 & 0.20 & c\\
55787 & $Y$ & VVV & 21.30 & 0.45 & p\\
55277 & $J$ & VVV & 20.81 & 0.25 & p\\
55277 & $H$ & VVV & 17.79 & 0.05 & c\\
55277 & $K_s$ & VVV & 15.80 & 0.02 & c\\
... & ... & ... & ... & ... & ... \\
\hline 
\multicolumn{2}{l}{Post-outburst}\\
\hline 
58183 & $W1$ & NEOWISE & 8.92 & 0.02 & c\\
58183 & $W2$ & NEOWISE & 7.37 & 0.01 & c\\
59351 & $J$ & SOFI & 14.57 & 0.06 & p\\
59351 & $H$ & SOFI & 12.51 & 0.05 & p\\
59351 & $K_s$ & SOFI & 11.05 & 0.03 & p\\
60024 & $K_s$ & IRSF & 14.20 & 0.10 & p\\
60783 & $K_s$ & IRSF & 13.94 & 0.01 & p\\
60783 & $H$ & IRSF & 16.04 & 0.01 & p\\
60794 & $J$ & FIRE & 18.48 & 0.05 & p\\
60820 & $K_s$ & REM & 13.99 & 0.05 & p\\

... & ... & ... & ... & ... & ... \\
\hline 
\hline 
\end{tabular}
\tablefoot{Infrared brightness of VVV-WIT-13. Data downloaded from online catalogues are marked as "c", whilst data extracted by aperture photometry in this work are marked with "p". The full version of the table is presented in the supplementary file.}
\label{tab:photometry}
\end{table}

\section{VIRAC photometry and PM of SPICY group G342.1+0.2}
\label{sec:AppA}

\begin{figure}
    \includegraphics[height=7cm]{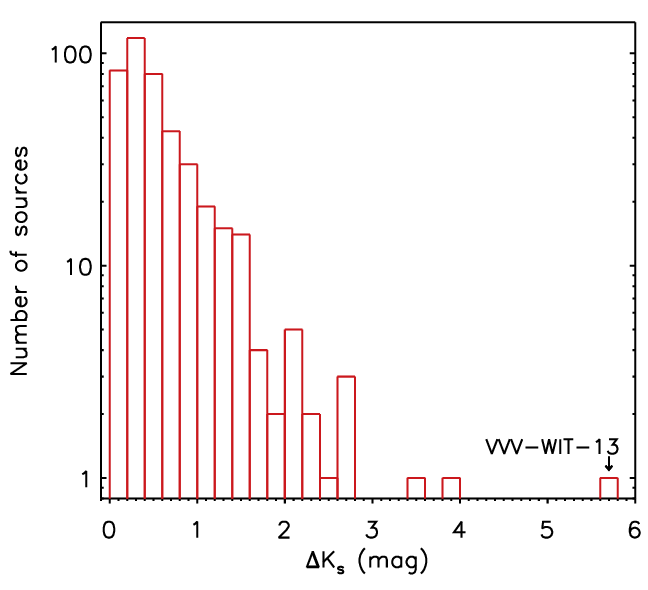}
        \includegraphics[height=7cm]{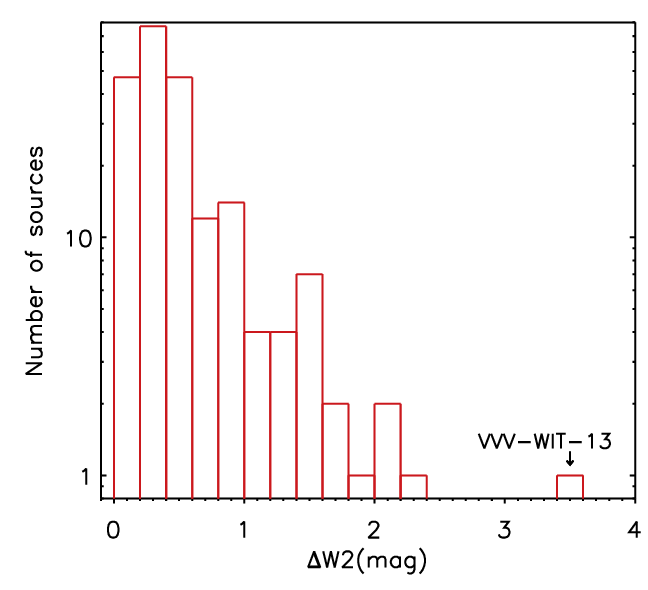}
      \caption{Histograms of $K_s$ and $W2$ amplitude of the members of SPICY group G342.1+0.2.}
         \label{fig: hist}
\end{figure}
   
Here we present the near- and mid-infrared variability of 442 YSO candidates in the SPICY group G342.1+0.2 by cross-matching the SPICY catalogue with VVV and NEOWISE time series. Overall, 422 sources have counterparts (distance less than 1 arcsec) in the VIRAC2 catalogue, and 220 sources have counterparts in the NEOWISE catalogue. 
The distribution of the $K_s$-band amplitude is presented in Figure \ref{fig: hist}, with 16\% of the sources having $\Delta K_s > 1$~mag and 6.6\% with $\Delta K_s > 2$~mag. Moreover, 19 out of 220 sources have $\Delta W1$ and $\Delta W2$ greater than 1~mag (see Figure \ref{fig: hist} for statistical plots).

We identified four objects that have eruptive light curve morphologies, including VVV-WIT-13. The light curves of the other three objects, SPICY 42942, SPICY 43289 and SPICY 43680 are presented in the left panels of Figure \ref{fig: LC_others}. Specifically, SPICY 42942 is classified as a Class II YSO in the SPICY catalogue, with $\Delta K_S = 3.56$~mag and $\Delta W2 = 0.83$~mag. Although its overall light curve morphology has an eruptive shape, there was a detection around MJD 55450 with $K_s = 14.15$ mag, which leads to confusion on the true variable nature of this source. We also checked the 2MASS $K_s$-band image \cite{Skrutskie2006} of SPICY 42942, and it was not seen on the image, indicating that it was fainter than 15.0~mag in $K_s$. Therefore, we still list it as a candidate eruptive YSO instead of a "dipper". SPICY 43289 exhibits typical eruptive light curves with $\Delta K_s = 3.84$~mag and $\Delta W2 = 1.71$~mag. The outburst lasted about 2500 to 3000 days based on the NEOWISE light curves.  SPICY 43680 had a low amplitude outburst with $\Delta K_s = 0.85$~mag and $\Delta W2 = 1.11$~mag. The duration of this event is on the scale of 1500 days. 

Eight sources exhibit long-term periodic variations that resemble mira-type variables (see Figure \ref{fig: LC_others}). The periods of these sources were extracted by a modified Lomb-Scargle method, which fits the light curve with a sinusoidal function plus a linear variation trend \citep[see][for more information]{Guo2022}. All eight variables have $\Delta K_s$ between 1 and 3 mag, which is 12\% of all group members within that amplitude range. The contamination should be noted when using the SPICY catalogue to identify YSOs. 

\begin{figure}
   \includegraphics[height=7.5cm]{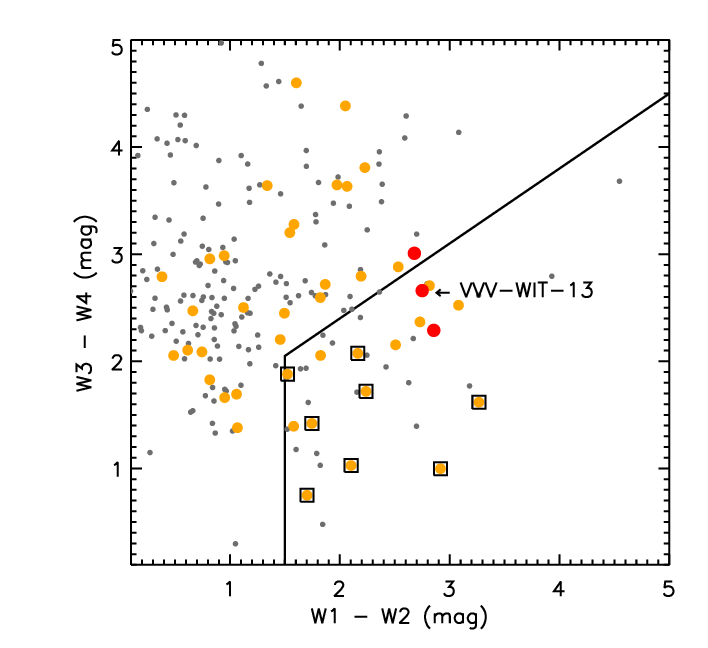}
     \caption{ALLWISE colour-colour diagram of sources from SPICY group G342.1+0.2. Data points are colour-coded by their VVV $K_s$ amplitude, as $ \Delta K_s < 1$ (grey), $1 \le \Delta K_s < 3$ (orange), and $ \Delta K_s \ge 3$ (red). Mira candidates are marked by squares. The solid lines are boundaries between regions more populated with YSOs (\textit{upper left}) and miras (\textit{bottom right}).}
\label{fig: WISE_CCD}
\end{figure}

\begin{figure*}
  \centering
     \includegraphics[height=9cm]{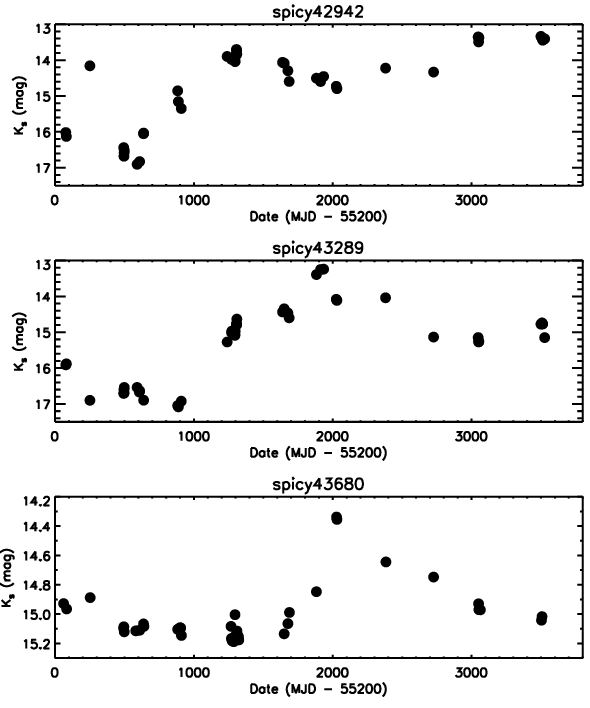}
     \includegraphics[height=9cm]{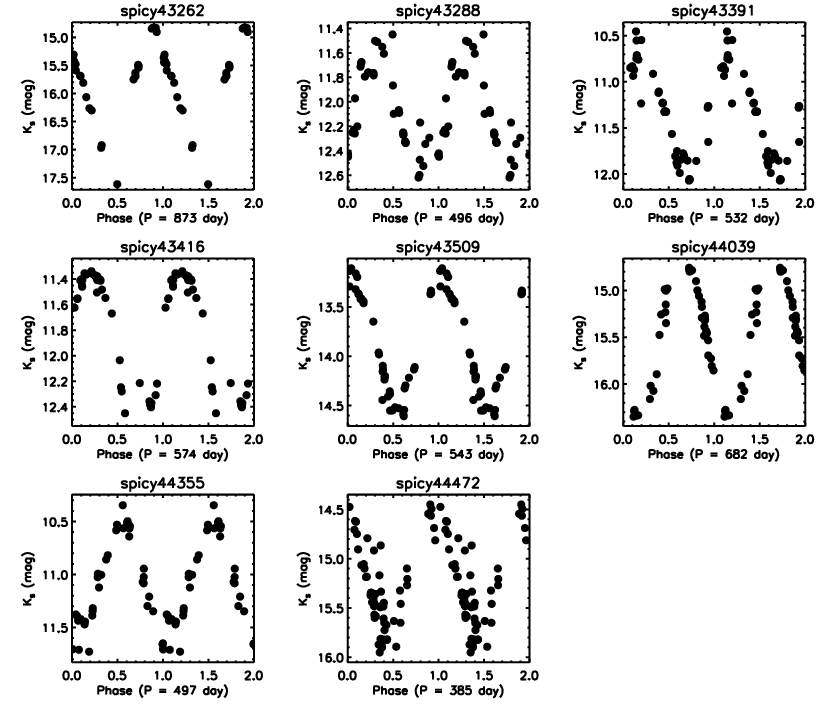}
      \caption{\textit{Left:} VVV light curves of three eruptive YSO candidates in the SPICY group G342.1+0.2. The names of each source are shown in the title of each plot. \textit{Right:} phase-folded VVV light curves of Mira candidates in the SPICY group G342.1+0.2, with periods presented on the x-axis.}
         \label{fig: LC_others}
   \end{figure*}

We obtained colour indices of the 219 YSO candidates from the SPICY group G342.1+0.2, with available detections in the ALLWISE catalogue. The $W1-W2$ and $W3-W4$ colour-colour diagram is presented in Figure \ref{fig: WISE_CCD}. The lower right region was drawn by \citet{Guo2022}, which is populated by mira-type variables. In the SPICY group G342.1+0.2, fourteen sources with $1 < \Delta K_s < 3$~mag are located in the region of Miras, including the eight Miras identified by their light curve morphologies. 

 We present the VIRAC PMs of the SPICY group G342.1+0.2. We found a median centre location at -1.3 and -2.8 mas/yr. The 1-sigma spread of the PMs is 2.35 mas/yr on the 2D-map. Notably, this spread is much larger than a typical young open cluster, suggesting SPICY group G342.1+0.2 is only a loose collection of stars instead of a dynamically bound group. The VIRAC parallax of VVV-WIT-13 has a large error bar $\varpi = 2.1 \pm 1.9$ mas/yr. Therefore, we did not use it to calculate the distance. In fact, most VIRAC parallaxes have similar large error bars.

     \begin{figure}
   \includegraphics[height=7.5cm]{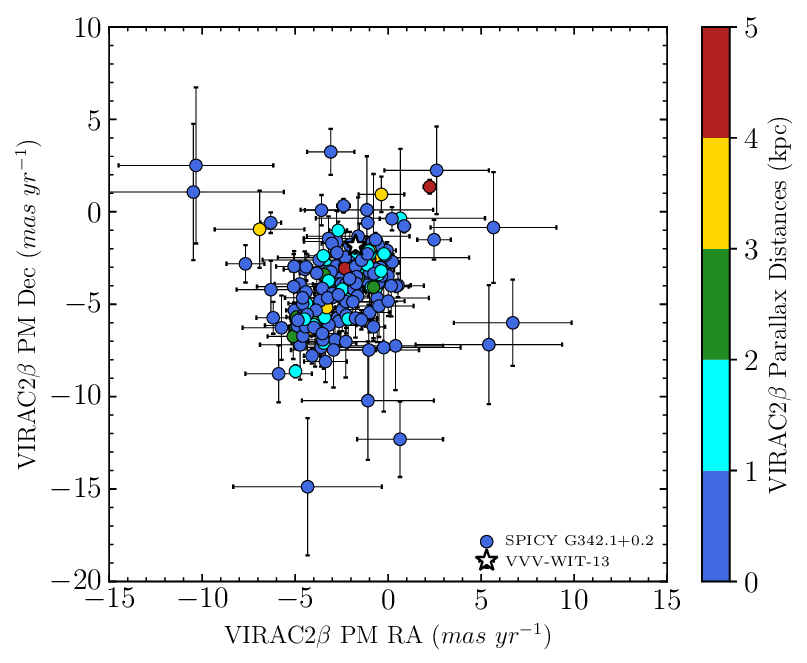}
      \caption{Proper motions of SPICY group G342.1+0.2 from VIRAC2$\beta$ catalog. VVV-WIT-13 is presented as the white star. The VIRAC parallax of this group is shown as the colour bar.}
   \label{fig: PM_virac}
   \end{figure}

\section{TiO band in optical spectrum}
\begin{figure}
   \centering
   \includegraphics[width=8cm]{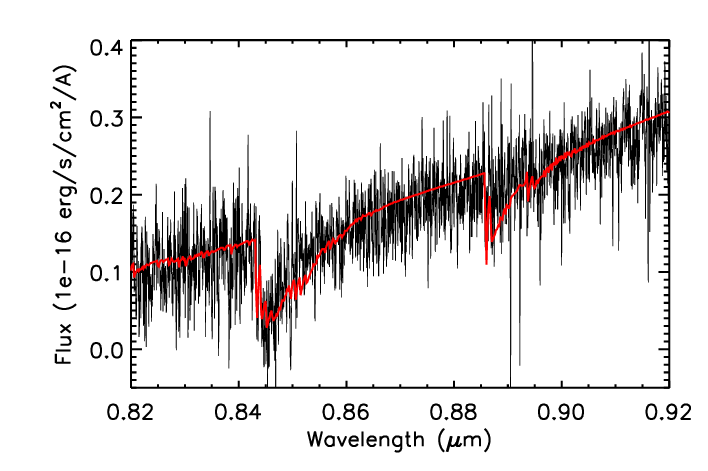}
      \caption{Optical spectrum from XSHOOTER, taken in the 2021 epoch. A TiO model with $T = 1300$~K is presented in red.}
         \label{fig: TiO}
   \end{figure}

The optical spectrum of VVV-WIT-13 was observed by XSHOOTER in the 2021 epoch. However, due to the high line-of-sight extinction, we only detected signals above 8000~\AA. Here, we present the spectrum in Figure \ref{fig: TiO}. Even with a low signal-to-noise ratio (S/N~$\sim$~4), we detected a TiO absorption band of the $\epsilon$ system near 8400~\AA. The bandhead of the $\delta$ system is also marked around 8800~\AA, but is not detected. We also present a rough fitting result using the spectral model from ExoMol, with a 1300~K temperature and RV~=~20~km/s. We note that due to the low S/N of the optical spectrum, we did not make analyses based on this result. However, we can confirm the existence of the TiO absorption band. 

\end{appendix}

\end{document}